\documentclass[
reprint,
 amsmath,amssymb,
 aps, physrev,
floatfix,
]{revtex4-2}

\usepackage{graphicx}
\usepackage{dcolumn}
\usepackage{bm}
\usepackage{booktabs}
\usepackage{siunitx}
\usepackage{hyperref}   
\usepackage{subcaption}
\usepackage{multirow}
\usepackage[english]{babel}
\usepackage{tabularx}
\usepackage{cleveref}   

\begin{document}


\title{On-axis top-up injection with multipole injection kicker in FCC-ee}

\author{S. Yue}
\author{Y. Dutheil}

    \affiliation{CERN, Geneva, Switzerland}

\author{M. Aiba}

    \affiliation{PSI, Villigen, Switzerland}

\date{\today}

\begin{abstract}
Top-up injection is required in FCC-ee to maintain high luminosity given the short beam lifetime.
The present baseline relies on on-axis off-energy injection with a conventional orbit bump; in this work, an alternative on-axis scheme based on a multipole injection kicker (MIK) is investigated.
The FCC-ee beam parameters place challenging demands on the MIK field distribution, which must combine a large field-free region for the circulating beam, a narrow transition region, and a sufficiently uniform field in the injection region across all operation modes.
To relax these challenging requirements while limiting the residual kick on the circulating beam, a compensated MIK scheme is proposed, in which a compensation MIK (CMIK) is placed upstream of the main MIK, and the required optics and field-symmetry conditions between the two devices are derived for different phase advances.
The pulsed-conductor kicker is adopted as the preferred design, and its influence on the circulating beam and its injection performance are evaluated for the Z operation mode. 
In the Z mode, the near-zero phase-advance compensated scheme limits the single-pass emittance growth of the circulating beam to about $7\%$ horizontally and below $2\%$ vertically, while the injected beam requires transfer-line pre-compensation to mitigate the field gradient in the injection region.
These results indicate that compensated MIK injection is a promising candidate for FCC-ee on-axis top-up injection.

\end{abstract}
 
\maketitle


\section{Introduction}
\label{sec:introduction}

The Future Circular lepton Collider (FCC-ee) is a proposed high-luminosity electron--positron collider designed to operate at several center-of-mass energies, with beam energies ranging from \SI{45.6}{GeV} at the Z pole to \SI{182.5}{GeV} at the $t\bar{t}$ threshold \cite{Benedikt:2928793}.
Because of the short beam lifetime expected under high luminosity operation, top-up operation is essential to maintain the beam current and maximize the integrated luminosity.

Several injection concepts have been studied for FCC-ee top-up injection~\cite{AIBA201898,DutheilFCCWeek2023TopupInjection}, covering both off-axis and on-axis off-energy injection, each of which can be realized with either a conventional local orbit bump or a multipole injection kicker (MIK).
On-axis off-energy injection is also referred to as synchrotron injection in the LEP context~\cite{CollierRoyEPAC1996PhysicsOpticsLEP,CollierPAC1995SynchrotronPhaseSpaceInjectionLEP}.
For all injection schemes in FCC-ee, injection is performed in the horizontal plane, where the dynamic aperture is larger than in the vertical plane \cite{FCCeeCDR2019}.

In off-axis injection, the injected beam undergoes horizontal betatron oscillations after injection.
These oscillations can generate a transverse offset near the interaction points (IPs), resulting in an enlarged synchrotron-radiation (SR) cone~\cite{AndreFCCWeek2023SRBackground}.
The larger SR cone can increase the load on the detector masks and may lead to additional detector background, especially because of the relatively slow transverse damping~\cite{CollierPAC1995SynchrotronPhaseSpaceInjectionLEP}.

In on-axis off-energy injection, the injected beam is placed on the chromatic closed orbit.
The separation from the circulating beam at the injection point is generated by the product of the horizontal dispersion and the injected beam energy offset.
Since the dispersion is designed to vanish at the IPs, the injected beam overlaps with the circulating beam in the detector regions, thereby reducing the additional SR background compared with off-axis injection.
On-axis off-energy injection was successfully used at LEP, where it demonstrated reduced background impact and high injection efficiency~\cite{CollierPAC1995SynchrotronPhaseSpaceInjectionLEP}.
For these reasons, on-axis off-energy injection is the favored top-up injection concept for FCC-ee~\cite{DutheilFCCWeek2023TopupInjection}.

The conventional implementation of top-up injection relies on a local orbit bump and a thin septum at the injection point.
This approach is mature and robust, and it has been widely applied in storage rings and colliders.
However, imperfect closure of the local orbit bump can excite residual betatron oscillations of the circulating beam.
Moreover, the septum thickness and the required mechanical clearance impose a lower limit on the separation between the circulating and injected beams.
These constraints become particularly important for on-axis off-energy injection when the momentum acceptance or the normalized dispersion is limited.
In addition, since the kicker acts directly on the circulating beam, a kicker failure can immediately lead to critical consequences, such as steering the circulating beam into the septum blade, which calls for dedicated machine-protection studies and mitigation measures~\cite{Ramjiawan2022FCCeeInjection, Dutheil2025FCCeeInjection, Hjelle2025FCCeeDamage,Yoshihara2025SBL}.

Considering these factors, an MIK provides a possible alternative.
An MIK generates a nonlinear magnetic field with a weak-field region around the circulating beam and a strong-field region at the injected beam position.
It can therefore provide the required deflection to the injected beam while reducing the perturbation to the circulating beam.
The transition between these two field regions performs a function analogous to that of a septum blade, but without introducing a physical material boundary between the two beams.
This motivates replacing the septum with an MIK in the baseline scheme, i.e., the orbit-bump injection (see Appendix).
The physical septum can instead be placed further upstream, where the thickness requirement is relaxed. 
Furthermore, this layout allows the baseline orbit-bump injection and the MIK injection to coexist as two complementary options.

MIK-based injection has been developed and demonstrated in several light sources.
The concept was first demonstrated at the KEK Photon Factory using a pulsed quadrupole magnet~\cite{PhysRevSTAB.10.123501}, and pulsed sextupole magnets were later adopted for top-up operation~\cite{PhysRevSTAB.13.020705}.
Pulsed-conductor schemes were subsequently proposed for BESSY~II~\cite{AtkinsonIPAC2011THPO024} and have since been studied or implemented in light sources such as MAX~IV~\cite{PhysRevSTAB.15.050705}, ALS-U~\cite{PhysRevAccelBeams.23.010702}, SOLEIL~\cite{PhysRevAccelBeams.26.020101}, ESRF-EBS~\cite{BenabderrahmaneIPAC2024TUPR47}, and ALBA~\cite{THPOPT047IPAC2022}, among others.

Table~\ref{tab:injection_schemes} summarizes injection concepts relevant to FCC-ee top-up injection.
The table is not intended to be an exhaustive survey, but rather to place the proposed scheme in the context of injection concepts that have been used or considered in storage rings and colliders.
In particular, on-axis off-energy injection has been used only in a limited number of colliders~\cite{CollierPAC1995SynchrotronPhaseSpaceInjectionLEP,Iida2026SuperKEKBSynchrotronInjection}, most notably LEP, while its combination with an MIK has not yet been demonstrated in an operating collider.
This gap motivates the present study of MIK-based on-axis off-energy injection as an alternative for FCC-ee top-up injection.

\begin{table*}[htbp]   
    \centering  
    \caption{Representative injection concepts for electron storage rings and lepton colliders, classified by injected-beam motion (rows) and kicker type (columns). Boldface marks the scheme studied here; the facility list is not exhaustive.}  
    \label{tab:injection_schemes}
    \begin{tabular}{lcc}
        \toprule
        & \textbf{Dipole kicker } & \textbf{Multipole injection kicker} \\
        & (Orbit bump / Short pulse kicker) & \\
        \midrule
        \textbf{Off-axis inj.}    & SuperKEKB, light sources, CEPC, \dots~\cite{Ohnishi2013SuperKEKB,AibaIPAC2018TopupReview,CEPCStudyGroup2024TDRAccelerator}  & KEK-PF, MAX IV, SOLEIL, \dots~\cite{PhysRevSTAB.13.020705,PhysRevSTAB.15.050705,PhysRevAccelBeams.26.020101} \\
        \addlinespace
        \textbf{On-axis inj.}     & LEP, FCC-ee baseline, SuperKEKB~\cite{CollierPAC1995SynchrotronPhaseSpaceInjectionLEP,DutheilFCCWeek2023TopupInjection,Iida2026SuperKEKBSynchrotronInjection}   &  \textbf{FCC-ee, this study}  \\
        \addlinespace
        \textbf{Swap-out inj.}   & APS-U, ALS-U, HEPS, CEPC, STCF, \dots~\cite{Calvey2025APSU,SteierIPAC2017WEPAB103,DuanIPAC2019TUPGW053,CEPCStudyGroup2024TDRAccelerator,Ai2025STCFCDRAccelerator}  & -- \\
        \bottomrule
    \end{tabular}
\end{table*}

Other advanced injection schemes have also been considered for storage rings, including swap-out, longitudinal, kick-and-cancel, and kickerless injection~\cite{AibaIPAC2018TopupReview,PhysRevSTAB.18.020701,PhysRevAccelBeams.28.060701}.
Swap-out injection, for instance, can relax the dynamic-aperture requirement by replacing a stored bunch with a freshly injected one, but it requires full-charge bunch injection, which is difficult for FCC-ee given the high bunch-charge requirements, booster limitations, and the potential impact on beam-beam performance.

Applying an MIK to FCC-ee is, however, more challenging than in typical light sources.
First, the momentum acceptance of the FCC-ee ring is limited to 1--2\%, while it can reach several percent in light sources.
Therefore, a large dispersion function at the MIK location is required, which tends to enlarge the beta function at the same location.
As discussed later, the large beta function helps reduce the effective septum thickness, whereas the corresponding large beam size is a significant disadvantage, tightening the MIK field requirements.
Second, the beam parameters also vary significantly among the Z, W, ZH, and $t\bar{t}$ operation modes, imposing different requirements on the field-free region, the transition region, and the injection region.
These features already make the MIK design for FCC-ee particularly demanding. 
Finally, stable FCC-ee operation requires minimizing the disturbance to the circulating beam during top-up injection.
To relax the MIK field requirements while limiting the residual kick on the circulating beam, we propose a compensated MIK scheme, in which a compensation MIK (CMIK) is installed upstream of the main MIK to cancel the residual kick.

The paper is organized as follows.
Section~\ref{sec:on_axis_requirements} establishes the FCC-ee on-axis injection requirements.
Section~\ref{sec:mik_topologies} compares representative MIK field topologies.
Section~\ref{sec:cmik_scheme} derives the optics and field-symmetry conditions for the compensation scheme.
Section~\ref{sec:near_zero_compensation} presents tracking studies of both the circulating and injected beams for the near-zero phase-advance layout. 
Section~\ref{sec:pi_phase_compensation} presents a proof-of-principle study of circulating-beam compensation for a $\pi$ phase advance between the CMIK and MIK.
Section~\ref{sec:mik_topology_discussion} discusses the prospects and limitations of different MIK topologies, and Sec.~\ref{sec:conclusion} summarizes the main conclusions.

\section{FCC-ee on-axis top-up injection requirements}
\label{sec:on_axis_requirements}

In on-axis off-energy injection, the injected beam is delivered with a relative momentum offset and is placed on the chromatic closed orbit of the collider.
At the injection point, the horizontal dispersion provides the transverse separation between the circulating and injected beams.
At the interaction points, where the dispersion is designed to be close to zero, the injected beam overlaps with the circulating beam, which reduces the additional synchrotron-radiation background compared with off-axis injection.

To achieve on-axis off-energy injection, the horizontal dispersion and the injected beam momentum offset must provide sufficient separation to clear both beam envelopes,
\begin{equation}
    |D_x \delta_{\rm inj}|
    \geq
    n_\sigma \sigma_{x,\rm cir} + S + n_\sigma \sigma_{x,\rm inj},
    \label{eq:on_axis_condition}
\end{equation}
where $D_x$ is the horizontal dispersion at the injection point, $\delta_{\rm inj}$ is the relative momentum offset of the injected beam, $\sigma_{x,\rm cir}$ and $\sigma_{x,\rm inj}$ are the horizontal beam sizes of the circulating and injected beams, $n_\sigma$ is the beam-envelope margin, and $S$ is the required clearance between the two $n_\sigma$ envelopes.
In this study, $n_\sigma=5$ is used for both beams.
The two remaining ingredients of Eq.~\eqref{eq:on_axis_condition}, namely the beam sizes $\sigma_x$ and the clearance $S$, are specified below.

The horizontal beam size is calculated as
\begin{equation}
    \sigma_x =
    \sqrt{\beta_x \epsilon_x + D_x^2 \sigma_\delta^2},
    \label{eq:sigma_x}
\end{equation}
with the corresponding beam parameters for either the circulating or the injected beam.
For the circulating beam, the collider dispersion and energy spread are included.
For the injected beam, matched betatron Twiss parameters are assumed at the injection point, while the dispersion contribution is neglected in the present estimate, so that $\sigma_{x,\rm inj}\simeq \sqrt{\beta_{x}\epsilon_{x,\rm inj}}$.
This corresponds to a dispersion-mismatched injected beam with respect to the collider closed orbit.
Such a mismatch increases the available separation at injection, but it can also excite betatron oscillations of the injected beam after injection.

Dividing both sides of Eq.~\eqref{eq:on_axis_condition} by $\sqrt{\beta_x}$ at the injection point makes the role of injection-point optics more explicit:
\begin{equation}
    \left|\frac{D_x}{\sqrt{\beta_{x}}} \delta_{\rm inj}\right|
    \geq
    (n_\sigma \sigma_{x,\rm cir} + S + n_\sigma \sigma_{x,\rm inj})/\sqrt{\beta_{x}}.
    \label{eq:on_axis_condition_nd}
\end{equation}

The left-hand side is now a product of normalized dispersion and the relative momentum offset.
The first term of the right-hand side is the normalized circulating beam size, which itself depends on the normalized dispersion.
Assuming a dispersion-free injected beam, the third term becomes $n_\sigma\sqrt{\epsilon_{x,\rm inj}}$, while the fixed clearance enters as $S/\sqrt{\beta_{x}}$.
In this form, the effective septum thickness decreases as the beta function at the injection point increases.  
The quantity $D_x/\sqrt{\beta_{x}}$ is therefore used as a figure of merit for the on-axis off-energy injection optics.

The clearance $S$ is set by the septum technology required for FCC-ee. 
Because the septum field must be maintained for up to one revolution period, corresponding to a flat-top of about $\SI{304}{\mu\second}$, magnetic-diffusion and skin-depth constraints prevent the septum blade from being made arbitrarily thin. 
Under the present septum design assumptions, the minimum blade thickness is \SI{2.8}{mm}, which is also adopted as the reference clearance for the transition region in the MIK study.

\begin{figure*}[htbp]
    \centering
    \includegraphics[width=\linewidth]{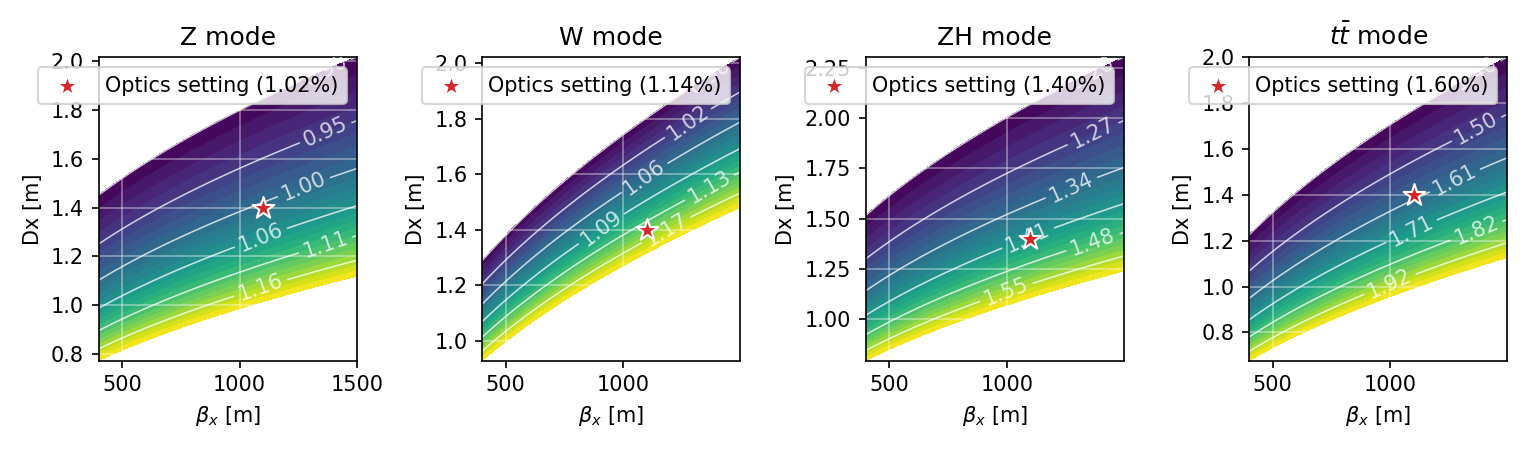}
    \caption{ 
     Horizontal dispersion $D_x$ required for on-axis off-energy injection as a function of the horizontal beta function $\beta_x$ at the injection point for the Z, W, ZH, and $t\bar{t}$ modes. 
     Contours give the required injected-beam momentum offset $|\delta_{inj}|$; red stars mark the selected optics. Both beams are taken at $5\sigma$ half-width with $S=\SI{2.8}{mm}$ separation. 
    }
    \label{fig:on_axis_injection_relation}
\end{figure*}

With the beam sizes and clearance fixed, Eq.~\eqref{eq:on_axis_condition} determines the required horizontal dispersion as a function of the beta function, the beam emittance, the energy spread, and the injected beam momentum offset.
Figure~\ref{fig:on_axis_injection_relation} shows the resulting $\beta_x$--$D_x$ relation at the injection point for the four FCC-ee operation modes and for several values of $\delta_{\rm inj}$.
For a fixed momentum offset, a larger beta function increases the betatron beam size and therefore requires a larger dispersion to maintain the same separation.

The present injection optics at the injection point has approximately
$D_x=\SI{-1.4}{m}$ and $\beta_x=\SI{1100}{m}$~\cite{LCC_106}, and the same injection optics is assumed for all four operation modes.
With $D_x$ fixed, Eq.~\eqref{eq:on_axis_condition} sets the minimum momentum offset needed in each mode: based on the corresponding beam parameters~\cite{Jebramcik2026LCC,Benedikt:2928793}, the required values are approximately $1.02\%$, $1.14\%$, $1.4\%$ and $1.6\%$ for the Z, W, ZH and $t\bar{t}$ modes, respectively.
The offset grows from the Z to the $t\bar{t}$ mode because the larger beam sizes demand a larger separation $|D_x\delta_{\rm inj}|$ at fixed dispersion.

The momentum offsets quoted above correspond to the minimum separation $S=\SI{2.8}{mm}$. 
The values listed in Table~\ref{tab:operation_modes} are larger in the Z and $t\bar{t}$ modes because those modes adopt a widened transition region ($S=\SI{3.6}{mm}$ and $\SI{4.5}{mm}$, respectively).

Since the required separation is obtained through the product $D_x\delta_{\rm inj}$, the momentum offset is effectively the free knob once the optics is chosen.
However, the usable offset is bounded by the available momentum acceptance.
If the required $\delta_{\rm inj}$ approaches this limit, a hybrid on-axis/off-axis injection scheme could be adopted~\cite{Yue:2912942,Benedikt:2928793}; in that case the residual betatron oscillation of the injected beam, especially near the IPs would need to be studied in detail.

Table~\ref{tab:operation_modes} summarizes the beam-envelope and field requirements that follow from these parameters~\cite{Jebramcik2026LCC,Chance2026HEB} and that are used in the MIK study.
The $\pm 5\sigma$ envelope of the circulating beam defines the field-free region, where the MIK field should be as small as possible; the $\pm 5\sigma$ envelope of the injected beam defines the injection region, where the MIK should provide a sufficient integrated field with good transverse uniformity; and the gap between the two envelopes is the field-transition region, whose nominal width is the reference clearance $S=\SI{2.8}{mm}$ introduced above.

Unlike typical light sources, FCC-ee has a large beam envelope at the injection point even with small emittance, because of the required large dispersion accompanied by a large beta function.
As the beam energy increases, the required field-free half-aperture grows from about \SI{9.6}{mm} in the Z mode to about \SI{13.3}{mm} in the $t\bar{t}$ mode, while the injected beam half-size grows from about \SI{1.8}{mm} to about \SI{6.2}{mm}.

The beam separation, i.e. the transition region in the MIK, could be widened, if additional momentum acceptance is available, or if a hybrid on-axis/off-axis scheme is adopted.
In this study, the beam separation is increased to \SI{3.6}{mm} in the Z mode, and to \SI{4.5}{mm} in the $t\bar{t}$ mode instead of the conservative \SI{2.8}{mm}, to obtain a smoother and more uniform field profile over the injected-beam footprint.
Consequently, the required injection location moves from about \SI{15}{mm} in the Z mode to about \SI{24}{mm} in the $t\bar{t}$ mode with respect to the reference orbit.
This large variation across operation modes demands substantial flexibility in the MIK field distribution.

Taking the required MIK deflection to be $\SI{100}{\mu\radian}$, the resulting integrated field (Table~\ref{tab:operation_modes}) remains comparable to values used in light-source MIK systems, even at the highest FCC-ee beam energy.
The main challenge is therefore not the field strength itself, but the simultaneous requirement of a large field-free region, a narrow transition region, and a sufficiently uniform injection field across all operation modes.

Finally, the $\pm 5\sigma$ envelope is a design convention and does not represent the full beam distribution.
Particles in non-Gaussian tails, or those affected by injection errors, may enter the field transition region and receive significantly larger kicks.
Dedicated halo, collimation, and machine-protection studies are therefore required before the final aperture and field-quality specifications can be defined.

\begin{table*}[htbp]
    \centering
    \caption{ 
    FCC-ee beam envelopes and MIK requirements at the injection point. Beam sizes are horizontal $5\sigma$ half-widths; separation is the gap between the two envelopes, and injection location is measured from the circulating-beam reference orbit. Boldface marks widened separations. Integrated-field values assume $\theta_x=\SI{100}{\mu\radian}$.}
    \label{tab:operation_modes}
    \begin{tabular}{lcccc}
        \toprule
        & \textbf{Z mode} & \textbf{W mode} & \textbf{ZH mode} & \textbf{$t\bar{t}$ mode} \\
        \midrule
        Beam energy [GeV]
        & 45.6 & 80 & 120 & 182.5 \\
        $5\sigma_{x,\rm cir}$ [mm]
        & 9.6 & 10.6 & 12.9 & 13.3 \\
        $5\sigma_{x,\rm inj}$ [mm]
        & 1.8 & 2.7 & 4.1 & 6.2 \\
        Beam separation [mm]
        & \textbf{3.6} & 2.8 & 2.8 & \textbf{4.5}\\
        Injection location [mm]
        & 15 & 16.1 & 19.8 & 24 \\
        Injected beam momentum offset [\%]
        & $1.07$ & $ 1.14$ & $ 1.4$ & $ 1.7$ \\
        $\int B_y\,\mathrm{d}s$ for $\theta_x = \SI{100}{\mu\radian}$ [mT\,m]
        & 15.2 & 26.8 & 40.0 & 60.8 \\
        \bottomrule
    \end{tabular}
\end{table*}

\section{Multipole injection kicker field concepts}
\label{sec:mik_topologies}

For collider top-up injection, a suitable MIK field must provide a sufficiently large field-free region for the circulating beam, a narrow field-transition region, and an injection region with small field gradient.
In addition, because the FCC-ee beam parameters vary substantially among operation modes, the field distribution should be flexible enough to cover all of them.
Two representative MIK field concepts are considered in this section: pulsed-multipole magnets and pulsed-conductor.


\subsection{Pulsed-multipole magnets}
\label{subsec:pulsed_multipole}

Pulsed-multipole magnets were first used for top-up injection at the KEK Photon Factory, where a pulsed sextupole magnet achieved good injection performance~\cite{PhysRevSTAB.13.020705}.
For FCC-ee, a single sextupole-like field is not sufficient to satisfy the required field-free region, transition region, and injection-field uniformity simultaneously.
Therefore, a combination of pulsed-multipole kickers, i.e., a series of individual pulsed multipoles installed close together, is considered in order to generate a more flexible nonlinear field profile.

For horizontal injection, the vertical beam size is small compared with the horizontal separation, so the field can first be studied along the horizontal axis.
In this conceptual design, quadrupole-, octupole-, and dodecapole-like components are used.
The vertical magnetic field on the horizontal axis is written as
\begin{equation}
    B_y(x) = C_1 x + C_3 x^3 + C_5 x^5 ,
    \label{eq:multipole_field}
\end{equation}
where $x$ is the horizontal position, and $C_1$, $C_3$, and $C_5$ are the strengths of the corresponding multipole components.
Adjusting these coefficients modifies the roots and gradients of the field profile, reducing the field in the circulating-beam region while providing the required deflection at the injected beam position. 
This makes the pulsed-multipole option relatively flexible for operation-mode changes.

\begin{figure}[htbp]
    \centering
    \begin{subfigure}[b]{0.4\textwidth}
        \centering
        \includegraphics[width=\linewidth]{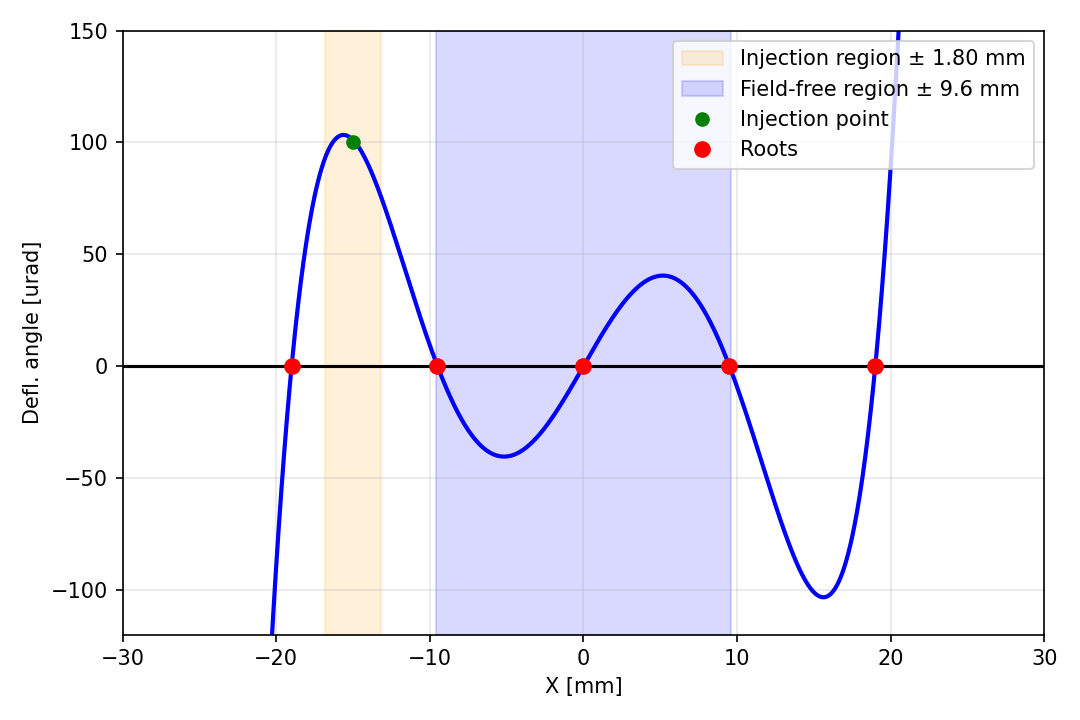}
        \caption{Z mode: $C_1=\SI{0.007}{\tesla\per\meter}$, $C_3=\SI{-129.3}{\tesla\per\meter\cubed}$, and $C_5=\SI{304352.7}{\tesla\per\meter\tothe{5}}$.}
        \label{fig:pulsed_multipole_z}
    \end{subfigure}
    \hfill
    \begin{subfigure}[b]{0.4\textwidth}
        \centering
        \includegraphics[width=\linewidth]{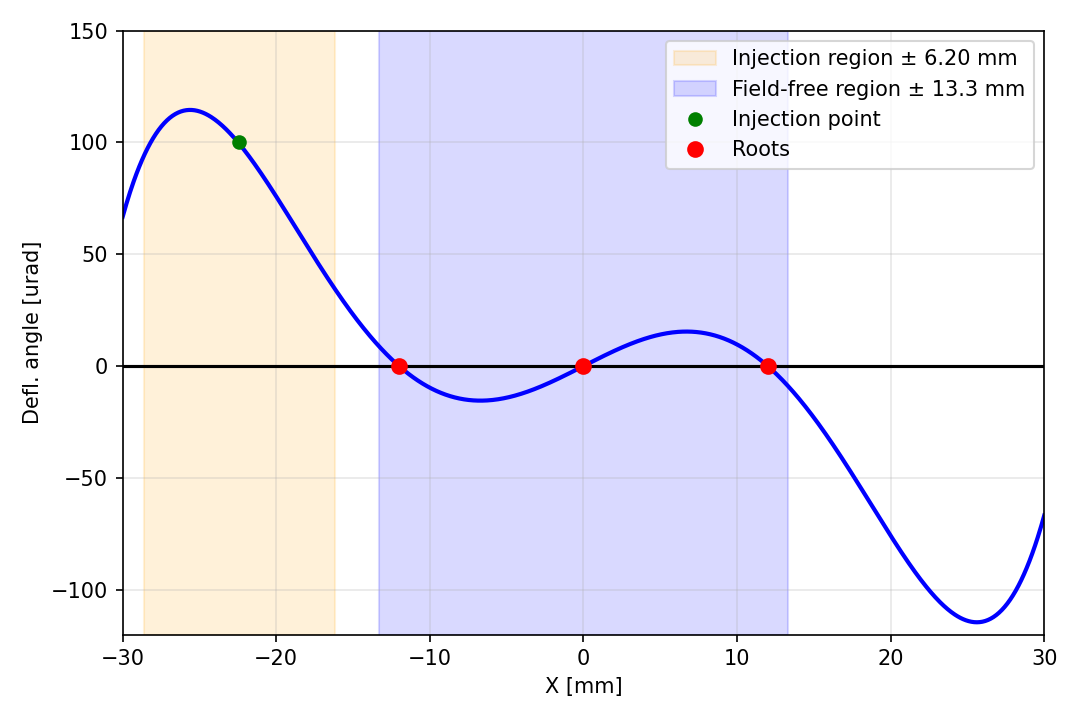}
        \caption{$t\bar{t}$ mode: $C_1=\SI{0.003}{\tesla\per\meter}$, $C_3=\SI{-27.7}{\tesla\per\meter\cubed}$, and $C_5=\SI{23725.6}{\tesla\per\meter\tothe{5}}$.}
        \label{fig:pulsed_multipole_ttbar}
    \end{subfigure}
    \caption{ 
    Horizontal deflection angle versus horizontal position for representative pulsed-multipole models in the Z (a) and $t\bar{t}$ (b) modes, with the coefficients listed below each panel and an effective length of \SI{1}{m}. Shading marks the field-free region (circulating beam) and the injection region; green dots are the injection point, red dots the field zero crossings.
    }
    \label{fig:pulsed_multipole}
\end{figure}

Figure~\ref{fig:pulsed_multipole} shows examples of the pulsed-multipole field distribution for the Z and $t\bar{t}$ modes; similar designs can also be obtained for the W and ZH modes.
The polynomial form allows the field-free region and injection region to be adjusted by changing the multipole strengths, which is attractive for FCC-ee because the beam envelopes vary significantly among operation modes.

However, the number of practically achievable multipole components is limited by magnet design and aperture constraints.
With a finite number of terms, an ideal field distribution is difficult to obtain over the full transverse range.
As a result, the residual field in the nominal field-free region is not negligible and can perturb the circulating beam during injection.
A compensation scheme is therefore introduced in Sec.~\ref{sec:cmik_scheme} to reduce this effect.

Another limitation is the narrow transition region between the circulating and injected beam envelopes.
With limited beam clearance, it is difficult to provide both the required deflection strength and a sufficiently small field gradient across the injected beam without increasing the residual field in the field-free region.
Therefore, transfer-line pre-compensation may be required to mitigate the effect of the MIK field gradient on the injected beam.

\subsection{Pulsed-conductor}
\label{subsec:pulsed_conductors}

The pulsed-conductor concept was proposed for BESSY~II and has since been studied or adopted in several light sources.
Compared with pulsed-multipole magnets, pulsed-conductor can provide a more favorable field profile, with a flatter field-free region around the circulating beam and a sufficiently strong field at the injected beam position.
However, its flexibility is more limited, because the field distribution is mainly determined by the conductor geometry, which cannot be changed during operation.

\begin{figure}[htbp]
    \centering
    \begin{subfigure}[b]{0.4\textwidth}
        \centering
        \includegraphics[width=\linewidth]{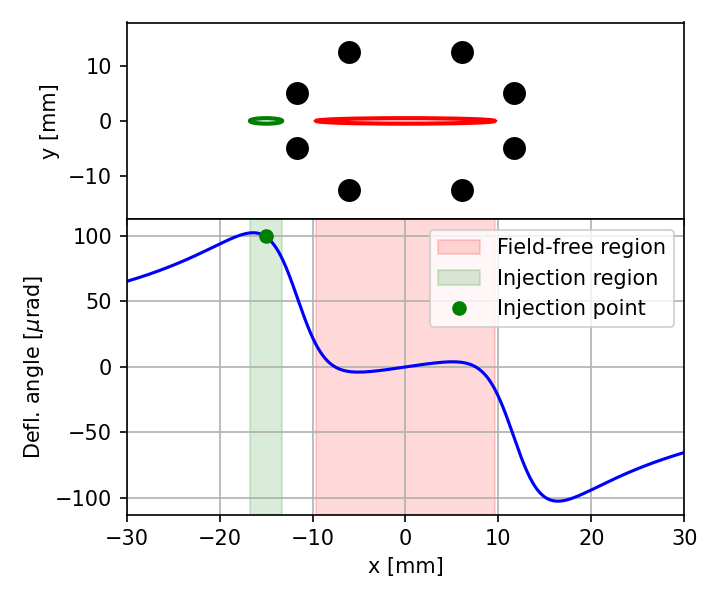}
        \caption{Z mode.}
        \label{fig:pulsed_conductor_z}
    \end{subfigure}
    \hfill
    \begin{subfigure}[b]{0.4\textwidth}
        \centering
        \includegraphics[width=\linewidth]{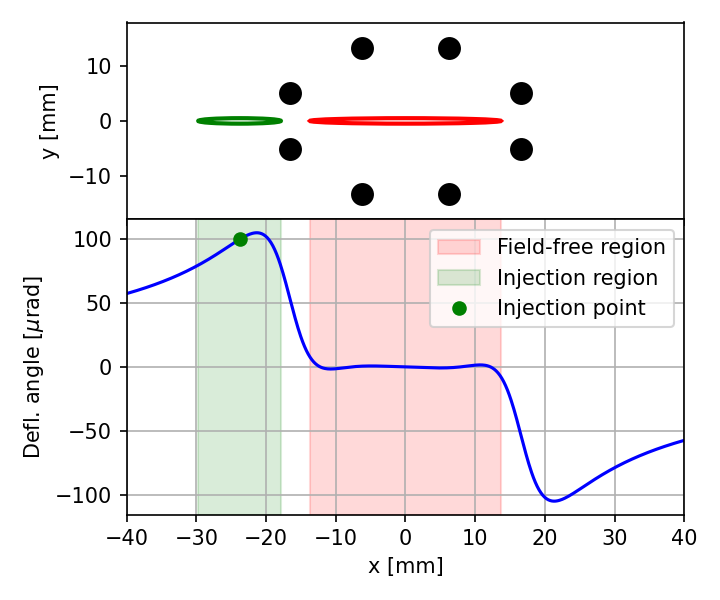}
        \caption{$t\bar{t}$ mode.}
        \label{fig:pulsed_conductor_ttbar}
    \end{subfigure}
    \caption{ 
   Representative pulsed-conductor models for the Z (a) and $t\bar{t}$ (b) modes. Upper panels: transverse layout of the eight \SI{1}{m}-long current elements (black) with the circulating (red) and injected (green) beams. Lower panels: horizontal deflection angle versus horizontal position, with the field-free and injection regions shaded and the injection point marked.}
    \label{fig:pulsed_conductor}
\end{figure}

Two representative pulsed-conductor layouts are considered, corresponding to the Z and $t\bar{t}$ modes, which have the smallest and the largest beam sizes, respectively.
Figure~\ref{fig:pulsed_conductor} shows the conductor locations and the corresponding field distributions for the two operation modes.
The conceptual layout uses eight conductors with left-right and up-down symmetry, all carrying the same current amplitude and direction.
The magnetic field is calculated using the Biot-Savart law for \SI{1}{m}-long current elements.
This simplified model is used to evaluate the field-distribution requirements and does not yet include detailed technical constraints such as conductor dimensions, cooling, insulation, mechanical supports, pulse-forming network limitations, or impedance effects.
 
The residual field in the field-free region is significantly reduced compared with the pulsed-multipole case, indicating a smaller perturbation to the circulating beam than in the pulsed-multipole magnets.

The narrow transition region is a common limitation of both MIK concepts, because it is difficult to obtain simultaneously a low residual field in the circulating-beam region and a sufficiently uniform field over the injected-beam footprint. 
With a transition-region width of \SI{2.8}{mm}, the field gradient across the injected beam remains appreciable; this is the quantitative reason for the widened transition regions adopted in the Z and $t\bar{t}$ modes in Sec.~\ref{sec:on_axis_requirements}.

According to Table~\ref{tab:operation_modes}, the beam requirements are similar for some operation modes.
It may therefore be possible to cover two or more modes with a single pulsed-conductor magnet by equipping it with additional conductors and adjusting the current amplitudes.
This requires a more detailed study of both the beam dynamics and the technical design, and is not pursued further in this conceptual study.

Despite its more limited flexibility, the pulsed-conductor model is adopted in the following tracking studies, because of its more favorable field distribution and, in particular, the lower residual field in the field-free region around the circulating beam.

\section{Compensated MIK scheme}
\label{sec:cmik_scheme}

For an ideal MIK field, the magnetic field should be negligible in the circulating beam region and sufficiently strong at the injected beam position.
In practice, however, the field in the nominal field-free region cannot be exactly zero.
The circulating beam can therefore receive a residual nonlinear kick during injection, especially given the large beam envelope of FCC-ee.
To reduce this perturbation, a compensated MIK scheme is considered.

\subsection{Optics requirements for the CMIK}
\label{subsec:cmik_optics}

The basic idea is to install a compensation MIK upstream of the main MIK.
The CMIK pre-distorts the circulating beam, and the main MIK cancels this distortion while providing the required kick to the injected beam.
For the circulating beam, the phase-space distribution before the CMIK and after the MIK should therefore be matched to the corresponding distribution without the MIK and CMIK fields.

\begin{figure}[htbp]
    \centering
    \includegraphics[width=\linewidth]{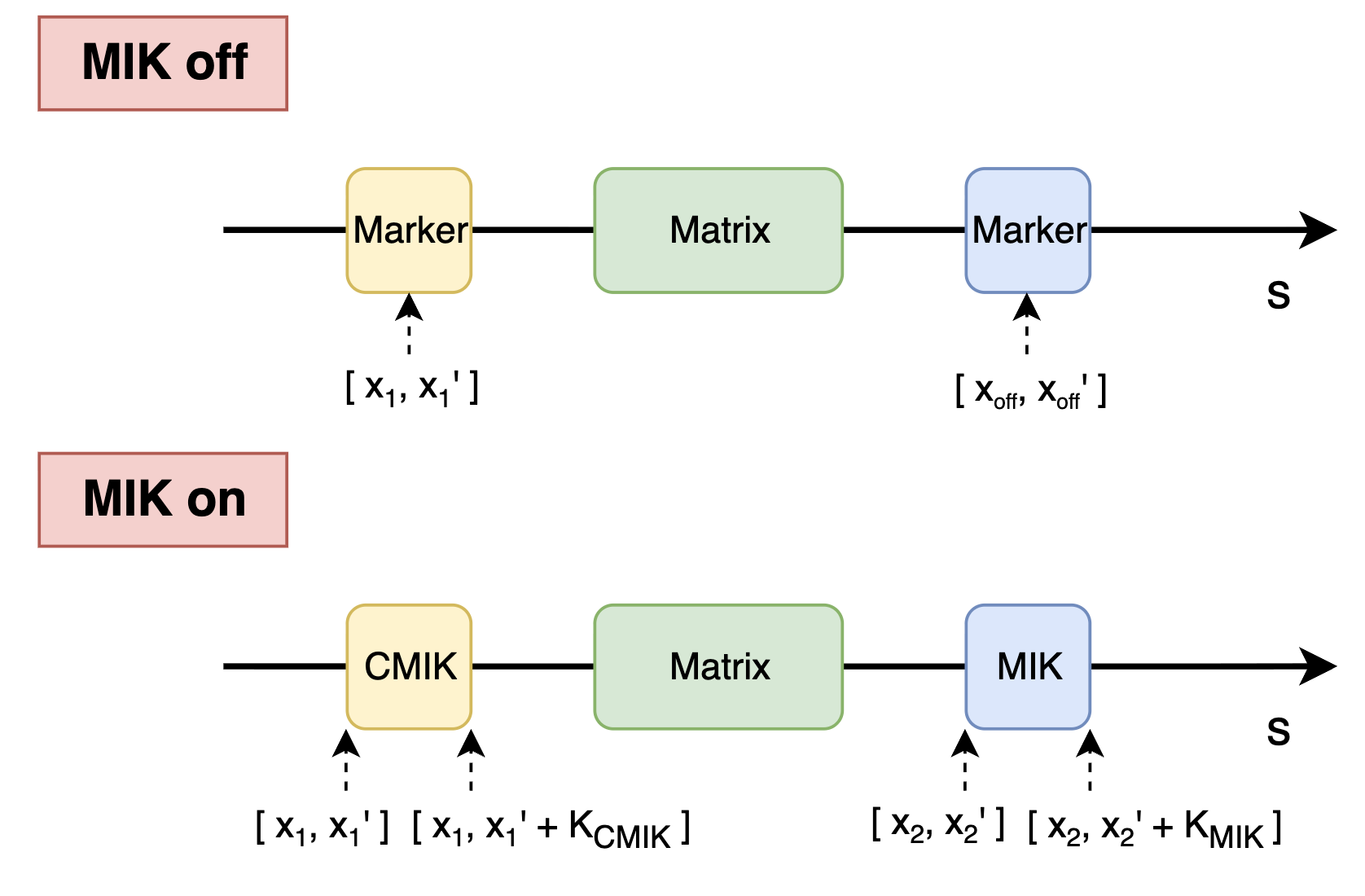}
    \caption{  
    Principle of the compensated MIK scheme for the circulating beam. Upper path: both devices off (reference). Lower path: the CMIK pre-distorts the beam, the distortion is transported by the linear map $M$, and the MIK cancels it.
    }
    \label{fig:compensation_sketch}
\end{figure}

Figure~\ref{fig:compensation_sketch} shows a schematic view of the compensation scheme.
The CMIK and MIK are separated by a transfer section represented by a transfer matrix.
When the MIK and CMIK are turned off, their locations are treated as markers.
Based on this layout, the conditions on the transfer matrix and the Twiss parameters at the CMIK and MIK required for compensation are derived in the following.

A simplified 5-D transfer matrix between the CMIK and MIK is used to derive the compensation conditions:
\begin{equation}
M =
\begin{bmatrix}
R_{11} & R_{12} & 0 & 0   & R_{16} \\
R_{21} & R_{22} & 0 & 0   & R_{26} \\
0 & 0 & R_{33} & R_{34}   & 0 \\
0 & 0 & R_{43} & R_{44}   & 0 \\ 
0 & 0 & 0 & 0   & 1
\end{bmatrix}.
\label{eq:cmik_matrix}
\end{equation}
Here, horizontal--vertical coupling and vertical dispersion are neglected.

Without the CMIK kick, the particle coordinates at the MIK location are
\begin{equation}
\begin{bmatrix}
x_{\rm off} \\
x'_{\rm off} \\
y_{\rm off} \\
y'_{\rm off} \\
\delta
\end{bmatrix}
=
M
\begin{bmatrix}
x_1 \\
x'_1 \\
y_1 \\
y'_1 \\ 
\delta
\end{bmatrix},
\label{eq:matrix_without_cmik}
\end{equation}
where the subscript ``off'' denotes the reference coordinates obtained when the CMIK is off.

With the CMIK turned on, the coordinates just before the MIK are
\begin{equation}
\begin{bmatrix}
x_2 \\
x'_2 \\
y_2 \\
y'_2 \\ 
\delta
\end{bmatrix}
=
M
\begin{bmatrix}
x_1 \\
x'_1 + K_{{\rm CMIK},x}(x_1,y_1) \\
y_1 \\
y'_1 + K_{{\rm CMIK},y}(x_1,y_1) \\ 
\delta
\end{bmatrix}.
\label{eq:matrix_with_cmik}
\end{equation}
The main MIK then applies the kicks $K_{{\rm MIK},x}(x_2,y_2)$ and $K_{{\rm MIK},y}(x_2,y_2)$ at the MIK location.
The compensation condition for the circulating beam is therefore
\begin{equation}
    x_2 = x_{\rm off},
    \qquad
    x'_2 + K_{{\rm MIK},x}(x_2,y_2) = x'_{\rm off},
\label{eq:compensation_condition_x}
\end{equation}
and
\begin{equation}
    y_2 = y_{\rm off},
    \qquad
    y'_2 + K_{{\rm MIK},y}(x_2,y_2) = y'_{\rm off}.
\label{eq:compensation_condition_y}
\end{equation}

Eqs.~\eqref{eq:matrix_without_cmik} and \eqref{eq:matrix_with_cmik} show that, for the position at the MIK to remain unchanged, the CMIK kick must satisfy
\begin{equation}
    R_{12}K_{{\rm CMIK},x}=0,
    \qquad
    R_{34}K_{{\rm CMIK},y}=0.
\end{equation}
For nonzero CMIK kicks, this leads to the phase-advance condition
\begin{equation}
    \sin\Delta\phi_x = 0,
    \qquad
    \sin\Delta\phi_y = 0.
\label{eq:phase_condition}
\end{equation}
Thus, the horizontal and vertical phase advances between the CMIK and MIK must be integer multiples of $\pi$.

Angular compensation requires
\begin{equation}
    R_{22}K_{{\rm CMIK},x}(x_1,y_1)
    +
    K_{{\rm MIK},x}(x_2,y_2)
    =0,
\label{eq:cmik_strength_x_general}
\end{equation}
and
\begin{equation}
    R_{44}K_{{\rm CMIK},y}(x_1,y_1)
    +
    K_{{\rm MIK},y}(x_2,y_2)
    =0.
\label{eq:cmik_strength_y_general}
\end{equation}
With the phase-advance condition of Eq.~\eqref{eq:phase_condition}, the diagonal transport terms reduce to
\begin{equation}
    R_{22}
    =
    \sqrt{\frac{\beta_{x1}}{\beta_{x2}}}\cos\Delta\phi_x,
    \qquad
    R_{44}
    =
    \sqrt{\frac{\beta_{y1}}{\beta_{y2}}}\cos\Delta\phi_y,
\end{equation}
where the indices 1 and 2 denote the CMIK and MIK locations, respectively.
The required strength relations are therefore
\begin{equation}
    \sqrt{\frac{\beta_{x1}}{\beta_{x2}}}
    \cos\Delta\phi_x\,
    K_{{\rm CMIK},x}(x_1,y_1)
    +
    K_{{\rm MIK},x}(x_2,y_2)
    =0,
\label{eq:cmik_strength_x}
\end{equation}
and
\begin{equation}
    \sqrt{\frac{\beta_{y1}}{\beta_{y2}}}
    \cos\Delta\phi_y\,
    K_{{\rm CMIK},y}(x_1,y_1)
    +
    K_{{\rm MIK},y}(x_2,y_2)
    =0.
\label{eq:cmik_strength_y}
\end{equation}

The field relation between the CMIK and MIK also depends on the coordinate transformation between the two locations.
From Eq.~\eqref{eq:matrix_with_cmik}, the horizontal coordinate at the MIK can be written as
\begin{equation}
    x_2
    =
    R_{11}x_1
    +
    R_{12}
    \left[
        x'_1 + K_{{\rm CMIK},x}(x_1,y_1)
    \right]
    +
    R_{16}\delta .
\end{equation}
With the phase-advance condition in Eq.~\eqref{eq:phase_condition}, $R_{12}=0$.
If $R_{16}=0$, the dependence on $\delta$ is removed and the coordinate relation becomes
\begin{equation}
    x_2
    =
    \sqrt{\frac{\beta_{x2}}{\beta_{x1}}}
    \cos\Delta\phi_x\, x_1 .
\label{eq:x_coordinate_scaling}
\end{equation}

$R_{16}=0$ requires the dispersion-matching condition
\begin{equation}
    D_{x2}
    =
    \sqrt{\frac{\beta_{x2}}{\beta_{x1}}}
    \cos\Delta\phi_x\, D_{x1},
\label{eq:dispersion_matching}
\end{equation}
where $D_{x1}$ and $D_{x2}$ are the horizontal dispersions at the CMIK and MIK, respectively.

This dispersion-matching condition is required because the MIK and CMIK fields act on the total horizontal position $x = x_\beta + D_x\delta$, not on the betatron coordinate alone.
Since the circulating beam has a finite energy spread, only when $D_{x2}$ satisfies this condition does the total coordinate scale by the same factor for all $\delta$, so that the field-symmetry relation between the CMIK and MIK holds across the whole energy distribution rather than only for on-momentum particles.

Combining Eqs.~\eqref{eq:cmik_strength_x} and \eqref{eq:x_coordinate_scaling}, the required CMIK-MIK strength relation in the horizontal plane can be written as

\begin{widetext}
\begin{equation}
    \sqrt{\frac{\beta_{x1}}{\beta_{x2}}}
    \cos\Delta\phi_x\,
    K_{{\rm CMIK},x}(x_1,y_1)
    +
    K_{{\rm MIK},x}
    \left(
        \sqrt{\frac{\beta_{x2}}{\beta_{x1}}}
        \cos\Delta\phi_x\,x_1,
        y_2
    \right)
    =0 .
\label{eq:scaled_compensation_x}
\end{equation}
\end{widetext}

The vertical relation is analogous.
Thus, the CMIK and MIK fields must satisfy both the strength relation and the coordinate-scaling relation imposed by the optics between the two devices.

If horizontal--vertical coupling is included in the transfer matrix, a kick in one plane can generate a position or angle change in the other plane.
Based on Eqs.~\eqref{eq:matrix_without_cmik} and \eqref{eq:matrix_with_cmik}, the additional variations at the MIK location can be written as
\begin{equation}
    \Delta x = R_{14}  K_{{\rm CMIK},y}(x_1,y_1), \quad \Delta x' = R_{24} K_{{\rm CMIK},y}(x_1,y_1)
\label{eq:coupling_x}
\end{equation}
and
\begin{equation}
    \Delta y =  R_{32}  K_{{\rm CMIK},x}(x_1,y_1), \quad \Delta y' = R_{42} K_{{\rm CMIK},x}(x_1,y_1)
\label{eq:coupling_y}
\end{equation}
These off-diagonal terms couple the two planes at the MIK: $R_{14}$ and $R_{24}$ feed the vertical CMIK kick into the horizontal plane, while $R_{32}$ and $R_{42}$ feed the horizontal CMIK kick into the vertical plane.

Similarly, the position-coupling terms $R_{13}$ and $R_{23}$ would add a vertical contribution to the horizontal coordinate at the MIK, and the vertical plane can be treated analogously.
Although compensation in the presence of coupling is in principle possible, it would require a more complicated field relation and transfer matrix.
An approximately uncoupled transfer section between the CMIK and MIK is therefore preferred.

With an uncoupled transfer section, the horizontal and vertical planes can be treated independently.
According to Eq.~\eqref{eq:phase_condition}, each phase advance must then be an integer multiple of $\pi$, but the two planes need not share the same value.
In the following, the CMIK-MIK strength relations are considered for phase advances of $0$, $\pi$, and $2\pi$ in each plane.
Note that a vanishing phase advance in one plane requires the CMIK and MIK to be co-located, so the phase advance in the other plane vanishes as well; this leaves the five independent combinations listed in Table~\ref{tab:cmik_mik_relation}.

For simplicity, the optics conditions in Table~\ref{tab:cmik_mik_relation} are chosen as
$\beta_{x1}=\beta_{x2}$, $\beta_{y1}=\beta_{y2}$, and
$D_{x2}=\cos\Delta\phi_x\,D_{x1}$.
The coordinate relations then reduce to
$x_2=\cos\Delta\phi_x\,x_1$ and
$y_2=\cos\Delta\phi_y\,y_1$.
Depending on the horizontal and vertical phase advances, the required CMIK and MIK kick functions are either equal or opposite in sign after the corresponding coordinate transformation.

Only normal multipole components are considered in this study. 
A normal multipole produces no vertical kick on the horizontal axis, $K_y(x,0)=0$, whereas a skew component would generate a vertical kick that increases with horizontal offset.
Skew components are therefore undesirable given the large horizontal extent of the circulating beam and are not considered further.
In the particular case $\Delta\phi_x=\Delta\phi_y=\pi$, the compensation conditions recover the optics requirement of the conventional $\pi$-bump: matched Twiss parameters at the CMIK and MIK separated by a betatron phase advance of $\pi$, with the two devices applying equal kicks of the same sign in the dipole limit.

\begin{table*}[htbp]
\centering
\caption{CMIK-MIK kick relations for different horizontal and vertical phase advances between the two devices, assuming uncoupled linear transport, equal beta functions at the two devices, and $D_{x2}=\cos\Delta\phi_x D_{x1}$. 
}
\label{tab:cmik_mik_relation}
\renewcommand{\arraystretch}{1.5}
\begin{tabular}{@{} c c l @{}}
\toprule
$\Delta\phi_x$ & $\Delta\phi_y$ & \multicolumn{1}{c}{Kick relation} \\
\midrule
$0$ & $0$ &
$\begin{aligned}
K_{{\rm MIK},x}(x,y) + K_{{\rm CMIK},x}(x,y) &= 0, \quad
K_{{\rm MIK},y}(x,y) + K_{{\rm CMIK},y}(x,y) = 0
\end{aligned}$
\\
\midrule
\multirow{2}{*}{$\pi$}
& $\pi$ &
$\begin{aligned}
K_{{\rm MIK},x}(x,y) - K_{{\rm CMIK},x}(-x,-y) &= 0, \quad
K_{{\rm MIK},y}(x,y) - K_{{\rm CMIK},y}(-x,-y) = 0
\end{aligned}$
\\
\cmidrule(l){2-3}
& $2\pi$ &
$\begin{aligned}
K_{{\rm MIK},x}(x,y) - K_{{\rm CMIK},x}(-x,y) &= 0, \quad
K_{{\rm MIK},y}(x,y) + K_{{\rm CMIK},y}(-x,y) = 0
\end{aligned}$
\\
\midrule
\multirow{2}{*}{$2\pi$}
& $\pi$ &
$\begin{aligned}
K_{{\rm MIK},x}(x,y) + K_{{\rm CMIK},x}(x,-y) &= 0, \quad
K_{{\rm MIK},y}(x,y) - K_{{\rm CMIK},y}(x,-y) = 0
\end{aligned}$
\\
\cmidrule(l){2-3}
& $2\pi$ &
$\begin{aligned}
K_{{\rm MIK},x}(x,y) + K_{{\rm CMIK},x}(x,y) &= 0, \quad
K_{{\rm MIK},y}(x,y) + K_{{\rm CMIK},y}(x,y) = 0
\end{aligned}$
\\
\bottomrule
\end{tabular}
\end{table*}

\section{CMIK-MIK scheme with near-zero phase advance}
\label{sec:near_zero_compensation}

The ideal compensation would place the CMIK and MIK at the same location, with a zero phase advance between them.
According to Table~\ref{tab:cmik_mik_relation}, the two devices then apply opposite kick functions, which cancel the residual kick on the circulating beam.
The injected beam, however, would pass through both devices at the same position, so the CMIK would also cancel the deflection from the MIK, leaving no net kick to place the injected beam on the chromatic closed orbit.
Exact zero-phase-advance compensation is therefore not realizable for injection.

A practical alternative is to keep the CMIK and MIK close together but at a finite longitudinal separation, so that the phase advance between them remains close to zero while leaving the space needed for injection.
The compensation is then only approximate, and the pulsed-conductor model (Sec.~\ref{sec:mik_topologies}) is used because a low residual field is important when the compensation cannot be exact.

\subsection{Injection-area layout}
\label{subsec:near_zero_layout}

\begin{figure*}[htbp]
    \centering
    \includegraphics[width=0.75\linewidth]{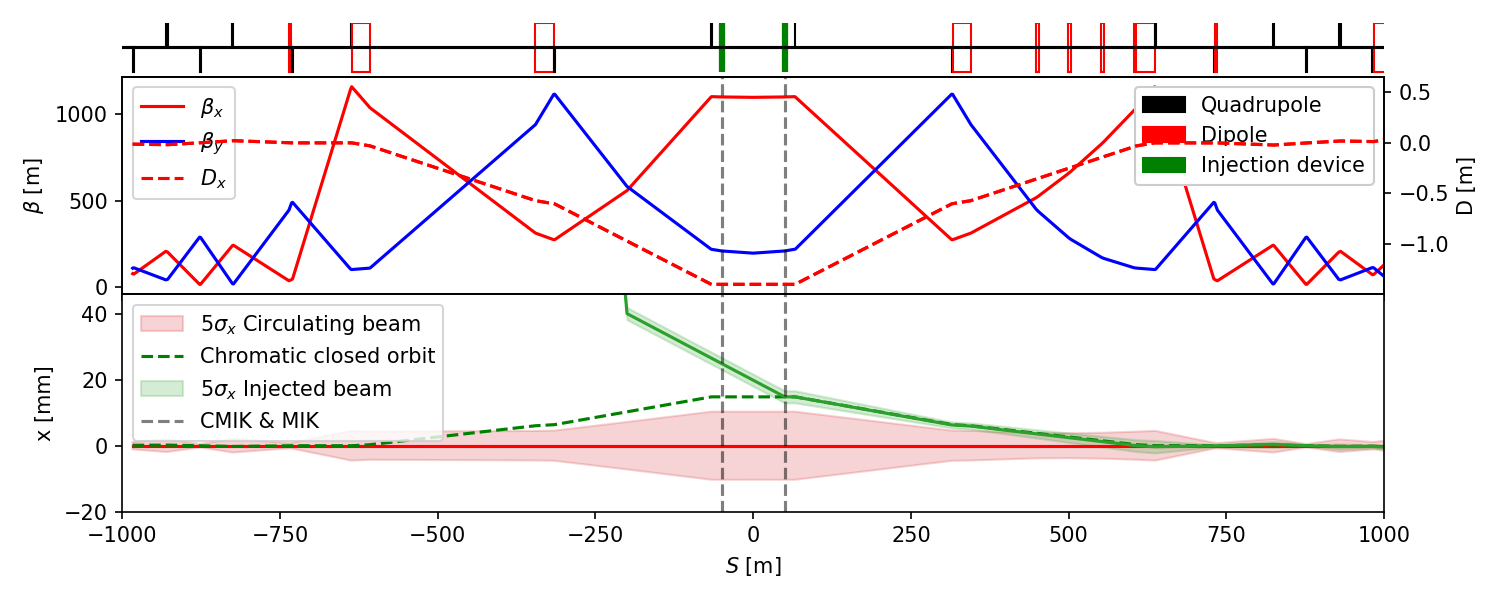}
    \caption{ 
    Near-zero phase-advance CMIK–MIK layout in the FCC-ee PB injection region, Z mode. Upper panel: lattice elements and optics functions $\beta_x$, $\beta_y$, $D_x$. Lower panel: $\pm 5 \sigma_{x}$ envelopes of the circulating and injected beams, chromatic closed orbit, and CMIK/MIK positions.
    }
    \label{fig:injection_layout_mu0}
\end{figure*}

In FCC-ee, the Point B (PB) straight section is dedicated to beam injection and extraction and has a length of approximately \SI{2}{km}. 
Figure~\ref{fig:injection_layout_mu0} shows the collider lattice and the beam layout in the PB section for the Z mode, based on the Local Chromaticity
Correction (LCC) lattice release~\texttt{LCC\_106.2.2}~\cite{LCC_106}.

The injection region is located near the middle of the straight section, where four bending magnets are used to generate and cancel the horizontal dispersion required for on-axis off-energy injection.

The available space for this system in the injection region is approximately \SI{130}{m}.
Within this space, the CMIK and MIK are installed with a longitudinal separation of \SI{100}{m}.
The optics parameters at the CMIK and MIK are nearly identical:
$\beta_{x,\rm MIK} \simeq \beta_{x,\rm CMIK} \simeq \SI{1100}{m}$,
$D_{x,\rm MIK} \simeq D_{x,\rm CMIK} \simeq \SI{-1.4}{m}$, and
$\beta_{y,\rm MIK} \simeq \beta_{y,\rm CMIK} \simeq \SI{210}{m}$.
The horizontal phase advance between the CMIK and MIK is about $0.01 \times 2\pi$, while the vertical phase advance is about $0.08 \times 2\pi$, owing to the smaller vertical beta function.

The lower panel of Fig.~\ref{fig:injection_layout_mu0} shows the circulating and injected beam trajectories during injection.
The circulating beam centroid should remain on the reference orbit when the MIK system is pulsed.
The injected beam is transported from the transfer line to the MIK, where it receives the required deflection and is sent to the chromatic closed orbit.
It subsequently merges with the circulating beam through synchrotron-radiation damping.

The CMIK and MIK must be pulsed for only one turn, to avoid kicking the injected beam on subsequent turns.
In the present layout, the CMIK is intended to act only on the circulating beam and not on the injected beam during injection.
Since the injected beam also traverses the CMIK region, a straightforward solution is to provide sufficient separation between the injected beam in the transfer line and the chromatic closed orbit, so that a field-shielding device can be installed in the CMIK region.

For a \SI{100}{m} separation between the CMIK and MIK, an MIK kick of $\SI{100}{\mu\radian}$ corresponds to a separation of approximately \SI{10}{mm} at the CMIK between the injected-beam trajectory and the chromatic closed orbit. 
If a larger beam-stay-clear margin is required, a larger MIK kick or a modified local layout could be considered.

It is also possible to let the injected beam pass through the CMIK without shielding.
Because the CMIK field is strongly position dependent, the field distribution could in principle be optimized such that the CMIK provides either zero kick or a controlled pre-kick to the injected beam before it reaches the MIK.
However, this introduces an additional constraint on the field relation between the CMIK at the pre-injection trajectory and the MIK at the final injection trajectory. 
Moreover, the CMIK field gradient across the injected beam would add further distortion. 
The present study assumes that the injected beam can be shielded from the CMIK field; the detailed design and integration of such a shielding device are beyond the scope of this paper.

\subsection{Circulating beam compensation in the Z mode}
\label{subsec:circulating_z_mu0}
  
The compensation performance is first evaluated for the circulating beam in the Z mode.  
Since the magnetic field depends on both transverse coordinates, the horizontal and vertical phase-space distributions are evaluated simultaneously. 
Tracking is performed using a matched Gaussian particle distribution with an rms relative energy spread of $0.12\%$ \cite{Jebramcik2026LCC}. 
At the CMIK, the corresponding $5\sigma$ beam sizes are \SI{9.6}{mm} in the horizontal plane and \SI{0.085}{mm} in the vertical plane.

\begin{figure}[htbp]
    \centering
    \includegraphics[width=0.95\linewidth]{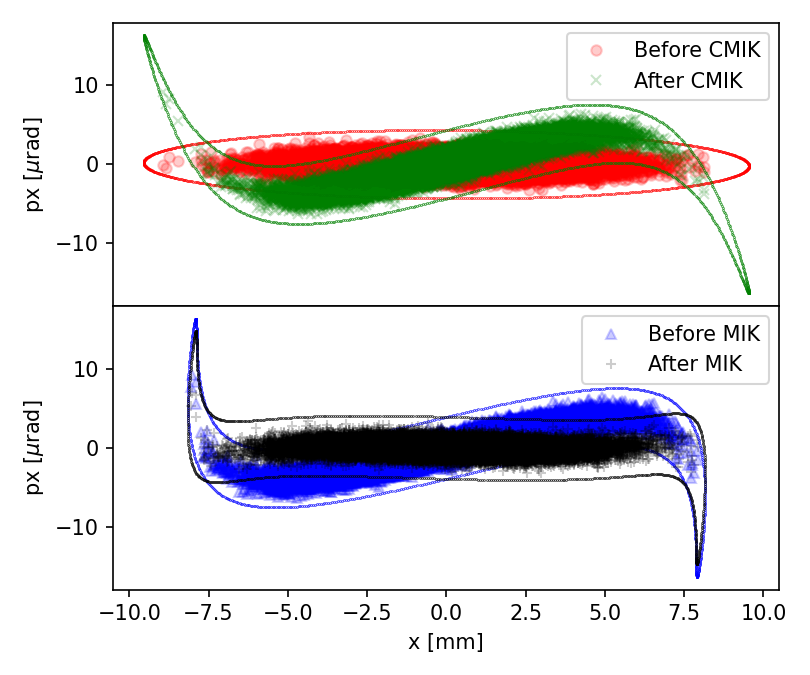}
    \caption{
    Single-pass horizontal phase space of the circulating beam in the Z mode near-zero phase-advance layout. 
    Upper: before (red) and after (green) the CMIK. 
    Lower: before (blue) and after (black) the MIK. Outlines are the corresponding $5\sigma$ contours.
    }
    \label{fig:MIK_hor}
\end{figure}

\begin{figure}[htbp]
    \centering
    \includegraphics[width=0.95\linewidth]{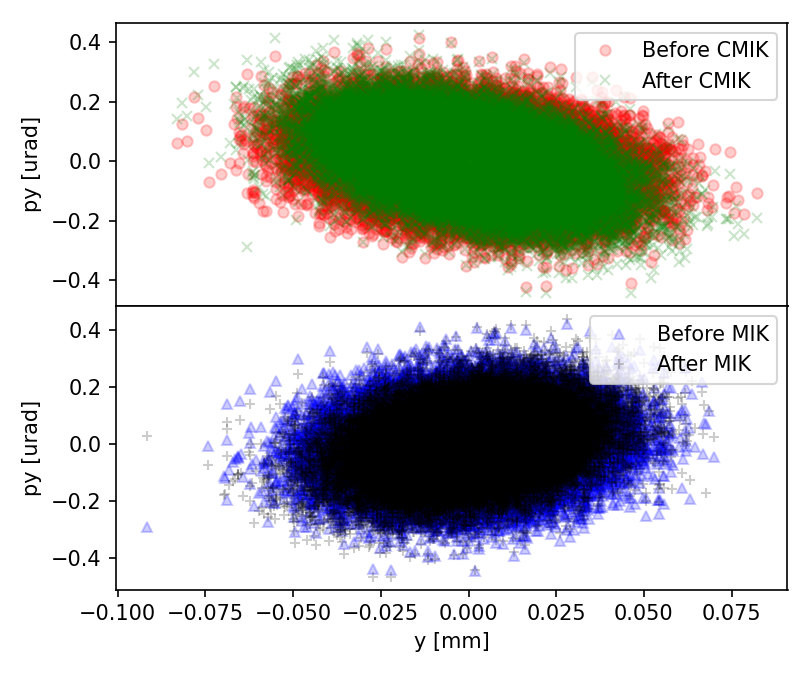}
    \caption{
    Single-pass vertical phase space of the circulating beam in the Z mode near-zero phase-advance layout. Upper: before (red) and after (green) the CMIK. Lower: before (blue) and after (black) the MIK.
    }
    \label{fig:MIK_ver}
\end{figure}

Figures~\ref{fig:MIK_hor} and \ref{fig:MIK_ver} show the single-pass evolution of the circulating beam in the horizontal and vertical phase spaces between the CMIK and MIK. After passing through the CMIK, the distribution is visibly distorted, indicating that the residual field of the pulsed-conductor model is not negligible; particles outside the nominal $5\sigma$ envelope receive larger kicks, because they enter the transition-field region where the gradient is stronger. After transport over the \SI{100}{m} CMIK-MIK separation, the distorted beam reaches the MIK, where the distortion is almost completely removed. Because the phase advance is close to, but not exactly, zero, particles near the edge of the distribution undergo a noticeable position change before the MIK, so the MIK cannot fully cancel the CMIK kick for every particle.

The geometric emittance is evaluated from the particle invariants with on-momentum Twiss parameters.
After the CMIK, the horizontal emittance increases by about a factor of four; after the MIK, the residual horizontal growth is reduced to approximately $7\%$. Because the vertical beam size is much smaller than the horizontal one, the vertical growth is less pronounced, remaining below about $2\%$. The near-zero phase-advance scheme therefore substantially mitigates the perturbation to the core of the circulating beam. Particles in the beam halo, however, can still receive significant deflections that are not fully compensated, so dedicated collimation and machine-protection studies are required for these uncontrolled particles.
The issue of halo particles is, however, not specific to the CMIK-MIK approach but common to most injection schemes.

\begin{figure}[htbp]
    \centering
    \includegraphics[width=\linewidth]{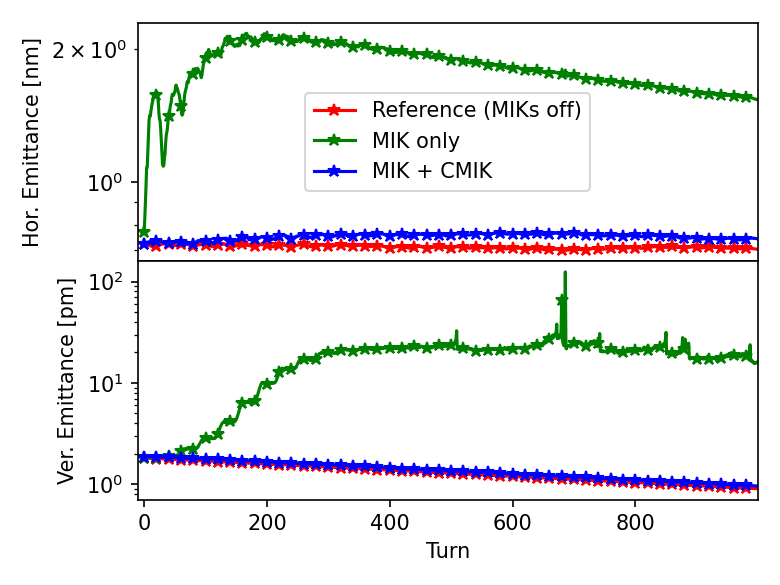}
    \caption{ 
    Multi-turn evolution of the horizontal (upper, nm) and vertical (lower, pm) geometric emittance of the circulating beam in the Z mode, on logarithmic scales: both MIKs off (red, reference), MIK only (green), CMIK + MIK (blue).
    }
    \label{fig:multiturn_emittance}
\end{figure}

The results above correspond to a single passage through the CMIK-MIK system.
To follow how the first-turn perturbation evolves over the subsequent turns, the circulating beam is also tracked over multiple turns.
Figure~\ref{fig:multiturn_emittance} compares the horizontal and vertical emittance evolution for three cases: (a) Reference: both MIKs off, (b) with the MIK only, and (c) with the CMIK-MIK compensation scheme. In this tracking, only synchrotron-radiation damping and quantum fluctuations~\cite{Xsuite} are included; lattice errors, beam-beam, and other collective effects are not included.
While this model is simplified, the goal here is to focus on the phase space perturbation caused by the CMIK and MIK. 

With the MIK alone, the residual field in the nominal field-free region drives a clear increase of both the horizontal and vertical emittance, followed by an apparent partial recovery as large-amplitude particles are lost. 
Once the CMIK is added, this growth is strongly suppressed: both planes follow the reference case closely and recover essentially the behavior obtained without the CMIK and MIK. 
This confirms, at the multi-turn level, that the compensation scheme keeps the perturbation to the circulating beam small, consistent with the single-pass emittance growth of about $7\%$ (horizontal) and below $2\%$ (vertical) found above.

\subsection{Injected beam transport in the Z mode}
\label{subsec:injected_z_mu0}
 
The injected beam is then studied in the same Z mode layout. 
In the present design, the injected beam from the transfer line passes through the \SI{100}{m} CMIK-MIK region without experiencing the CMIK field. 
It reaches the MIK at a horizontal offset of about \SI{15}{mm} from the reference orbit, where the MIK provides the nominal kick of $\SI{100}{\mu\radian}$.

At the injected beam position, the pulsed-conductor field exhibits a clear gradient across the beam, as shown in Fig.~\ref{fig:pulsed_conductor_z}. 
This field gradient contains not only a quadrupole component but also higher-order multipole components, which distort the injected-beam phase-space distribution. 
Since the distortion cannot be corrected in the collider ring, we propose to pre-deform the injected beam in the transfer line, so that the deformation is canceled by the MIK field when the beam reaches the MIK.

\begin{figure*}[htbp]
    \centering
    \includegraphics[width=0.8\linewidth]{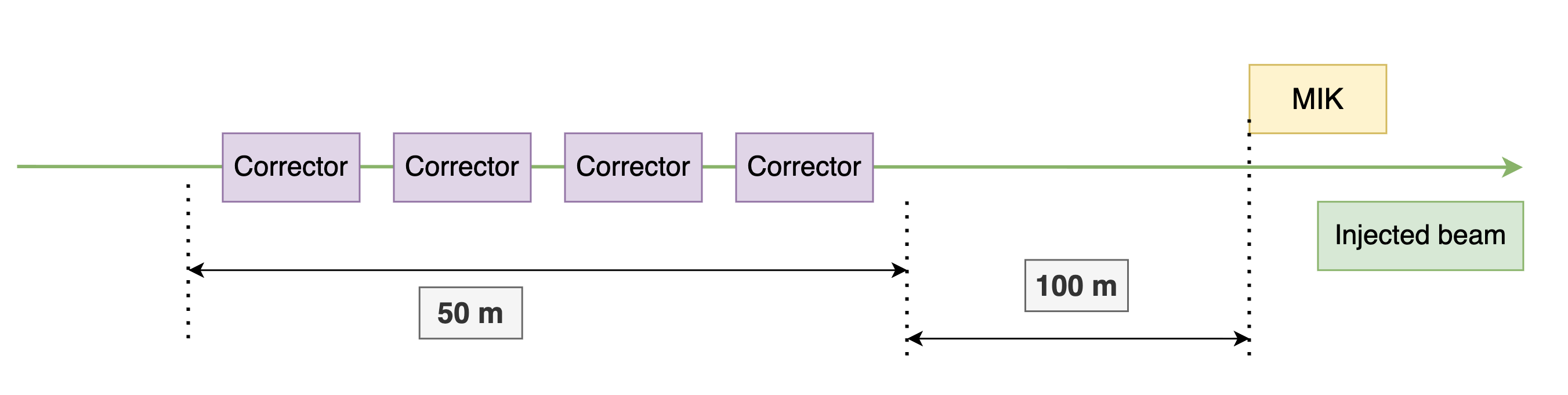}
    \caption{ 
    Conceptual transfer-line pre-compensation layout for the injected beam in the Z mode: four combined-function nonlinear correctors spanning \SI{50}{m}, ending \SI{100}{m} upstream of the MIK.
    }
    \label{fig:tl_compensation}
\end{figure*}

As a preliminary concept, four combined-function nonlinear corrector elements are installed in the upstream transfer line, each providing independently adjustable quadrupole and sextupole components.
As shown in Fig.~\ref{fig:tl_compensation}, approximately \SI{50}{m} is available for these nonlinear correctors.
The injected beam distribution can be optimized by adjusting the strengths of the four nonlinear correctors together with initial Twiss parameters, so as to maximize the fraction of the beam accepted by the collider ring.

For this preliminary study, $3\sigma$ and $5\sigma$ injected beam ellipses are used to characterize the distribution. 
The injected beam emittance is taken as \SI{0.12}{nm} in the horizontal plane and \SI{10}{pm} in the vertical plane.

\begin{figure*}[htbp]
  \includegraphics[width=0.95\linewidth]{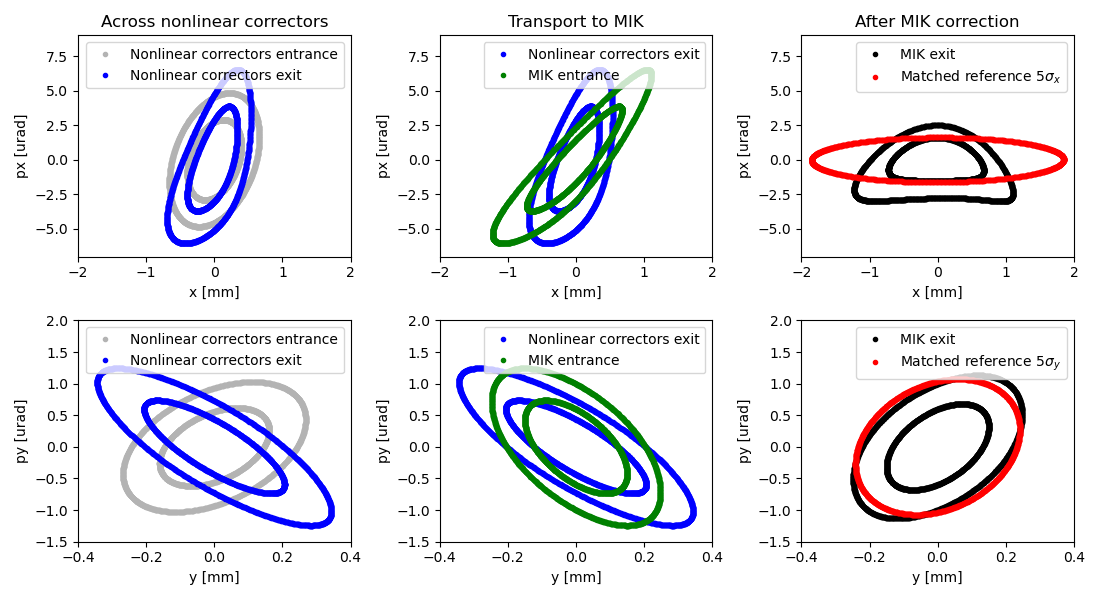}
    \caption{ 
    Injected-beam phase-space contours (inner $3\sigma$, outer $5\sigma$) in the Z mode; upper row horizontal, lower row vertical. Columns: across the nonlinear correctors (entrance grey, exit blue); transport to the MIK (corrector exit blue, MIK entrance green); after the MIK (black) against the $5\sigma$ orbit-bump reference acceptance (red). The nominal $\SI{100}{\mu\radian}$ MIK kick is subtracted.
}
    \label{fig:injection_compensation}
\end{figure*}

Figure~\ref{fig:injection_compensation} shows the injected beam ellipse distribution upstream and downstream of the transfer line nonlinear correctors, and before and after the MIK, in both horizontal and vertical phase space. 
The nominal $\SI{100}{\mu\radian}$ kick of the MIK is subtracted in Fig.~\ref{fig:injection_compensation}, so that the distortion before and after the MIK can be compared directly.
The red ellipse represents the $5\sigma$ injected beam envelope obtained for the reference orbit-bump injection, and it is used here as a provisional collider acceptance target.

The initial Twiss parameters upstream of the nonlinear correctors are: $\beta_x = \SI{145}{m}$, $\beta_y = \SI{287}{m}$, $\alpha_x = -0.36 $ and $\alpha_y = -0.46$.
After passage through the MIK, the horizontal $3\sigma$  ellipse lies within the provisional acceptance contour defined by the reference case. Particles between the $3\sigma$  and $5\sigma$  contours may be lost if the available dynamic aperture or momentum acceptance is limited, but their impact on the overall injection efficiency is expected to be small. The vertical phase-space distribution is largely preserved. Taken together, the horizontal and vertical results indicate that the transfer-line pre-compensation substantially mitigates the phase-space distortion induced by the MIK field gradient.

\begin{figure*}[htbp] 
    \centering
    \includegraphics[width=\linewidth]{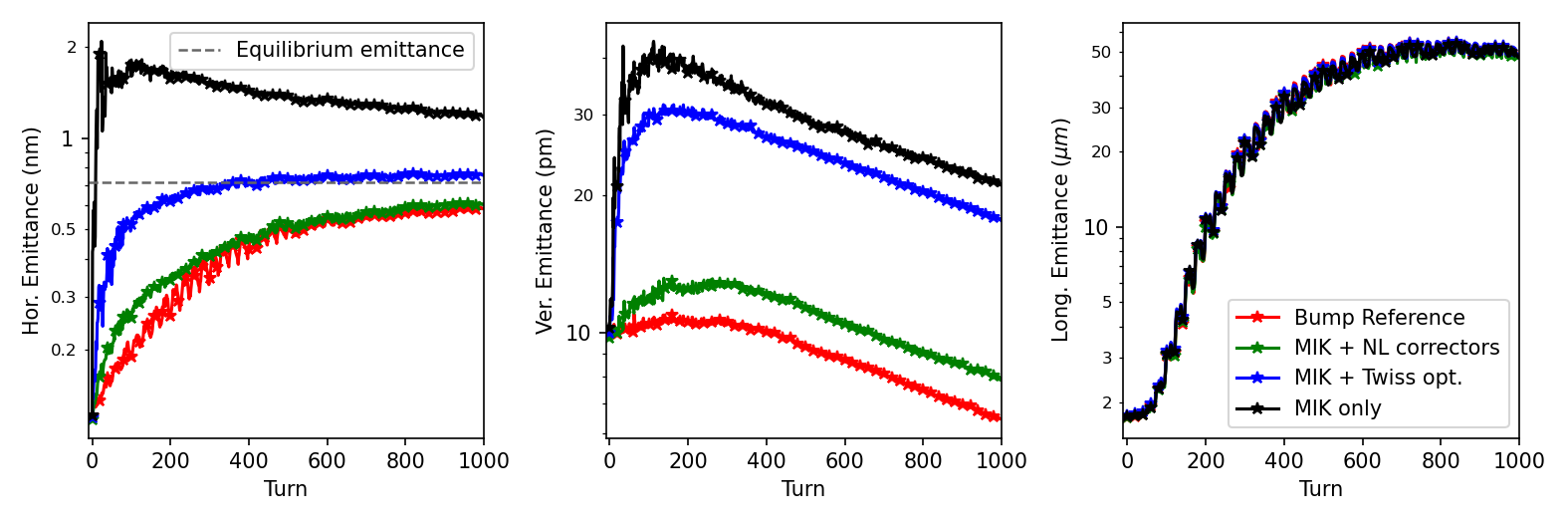}
    \caption{
    Multi-turn evolution of the horizontal (left, nm), vertical (middle, pm) and longitudinal (right, $\mu m$) emittance of the injected beam in the Z mode, on logarithmic scales: orbit-bump injection (red, reference), MIK with nonlinear correctors (green), MIK with Twiss optimization only (blue), MIK only without optimization (black).
    }
    \label{fig:injected_emittance_evolution}
\end{figure*}

Multi-turn tracking of the injected beam is then performed including synchrotron radiation damping and quantum fluctuations~\cite{Xsuite}, using 5000 particles followed over 1000 turns, with the MIK pulsed on the injection turn only.
Figure~\ref{fig:injected_emittance_evolution} compares the emittance evolution for four cases:
\begin{itemize}
    \item \textit{Bump injection}: the conventional orbit-bump scheme, in which the beam is
          injected through a septum rather than through the MIK; this case serves as the reference.
    \item \textit{MIK with nonlinear correctors}: the four transfer-line nonlinear correctors of
          Fig.~\ref{fig:tl_compensation} are used to pre-compensate the injection beam.
    \item \textit{MIK with Twiss optimization only}: only the injected beam Twiss parameters are optimized, without
          nonlinear correctors.
    \item \textit{MIK only without optimization}: no transfer line pre-compensation is applied, and the injected beam is matched directly to the off-momentum Twiss parameters of the collider lattice.
\end{itemize}
The emittance evolution obtained with the nonlinear correctors is very close to that of the reference bump injection, which shows that the transfer-line pre-compensation effectively mitigates the impact of the MIK field gradient. Optimizing the Twiss parameters alone compensates the dominant linear component, but the higher-order components remain: an emittance blow-up is still observed, most visibly in the vertical plane.

Without any compensation, the injected beam is strongly deformed and suffers significant losses. 
The Twiss-optimization option is therefore attractive for the Z mode, but it is not expected to be sufficient in the $t\bar{t}$ mode, where the higher-order components of the field gradient are considerably stronger (Fig.~\ref{fig:pulsed_conductor_ttbar}); the transfer-line nonlinear correctors are consequently the preferred solution.

\begin{figure}[htbp]
    \centering
    \includegraphics[width=0.9\linewidth]{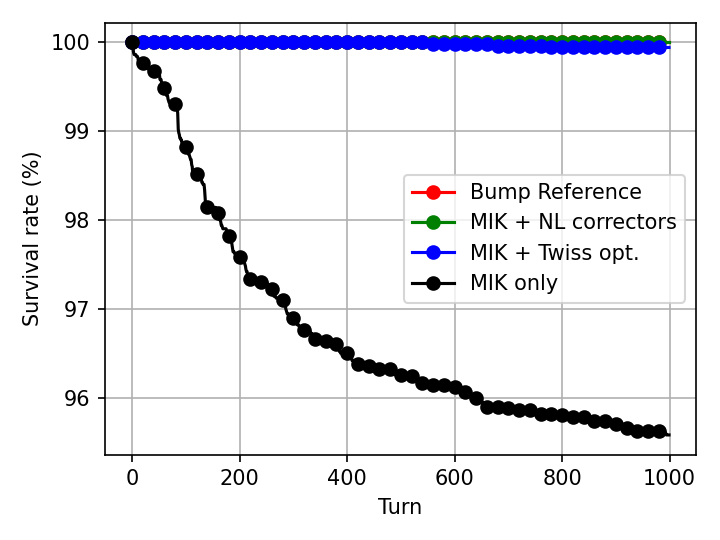}
    \caption{
    Survival rate of the injected particles versus turn number in the Z mode: orbit-bump injection (red, reference), MIK with nonlinear correctors (green), MIK with Twiss optimization only (blue), and MIK only without optimization (black). The first two cases overlap at $100\%$.
    }
    \label{fig:injection_efficiency}
\end{figure}

Figure~\ref{fig:injection_efficiency} shows the corresponding survival rate of the injected particles as a function of turn number. 
No particle loss is observed over 1000 turns either for the reference bump injection or for the MIK injection with transfer-line nonlinear correctors, whereas the Twiss-optimized case retains about $99\%$. 
Without any compensation the survival rate is still above $95\%$ after 1000 turns, but further losses are expected over the following few thousand turns.

The tracking uses 5000 particles with an initial Gaussian distribution and does not include beam-beam effects, lattice imperfections, or collimation, so it does not fully describe the behavior of the beam tails near $5\sigma$; it nevertheless supports the benefit of the transfer-line pre-compensation.
Further optimization of both the transfer line nonlinear correctors configuration and the MIK field uniformity will be pursued for the Z mode and the other operation modes.

\section{CMIK-MIK scheme with \texorpdfstring{$\pi$}{pi} phase advance}
\label{sec:pi_phase_compensation}

Another possible compensation scheme is to place the CMIK and MIK with horizontal and vertical phase advance of $\pi$.
Compared with the near-zero phase-advance scheme, this configuration can in principle provide a more exact compensation of the circulating beam, but it requires a dedicated redesign of the injection optics.

\begin{figure*}[htbp]
    \centering
    \includegraphics[width=0.75\linewidth]{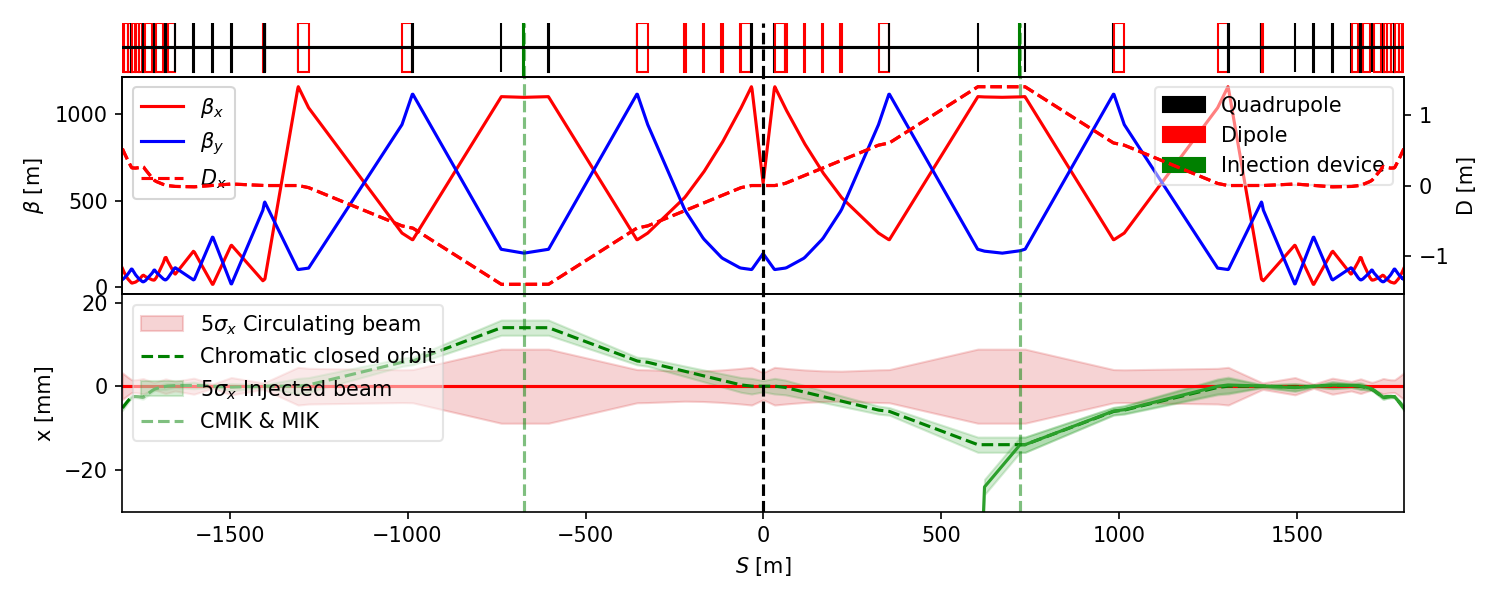}
    \caption{  
    Conceptual Z mode CMIK–MIK layout with horizontal and vertical phase advances of $\pi$ in the FCC-ee PB injection region. Upper panel: optics functions and device locations. Lower panel: $\pm 5\sigma_x$ envelopes of the circulating and injected beams and the chromatic closed orbit.
    }
    \label{fig:injection_layout_piphase}
\end{figure*}

Based on the PB injection optics, a mirrored injection lattice is constructed to test the compensation principle under $\pi$ phase advance conditions, as shown in Fig.~\ref{fig:injection_layout_piphase}.
To satisfy the compensation conditions, the horizontal and vertical beta functions at the CMIK and MIK are matched to be equal, while the horizontal dispersions at the two locations are made equal in magnitude and opposite in sign.
A \texttt{LineSegmentMap}, which represents a simplified beamline segment~\cite{Xsuite}, is inserted at the center of the injection region to match the two halves of the lattice, and the polarities of four bending magnets are reversed to obtain the required dispersion relation. 
This should be regarded as a conceptual transfer-map model rather than a full lattice design, and a dedicated redesign would be required before this option could be adopted for FCC-ee.

In this layout, the injected beam is guided into the chromatic closed orbit by the MIK.
Since the CMIK is far upstream of the MIK, the injected beam can in principle be separated from the CMIK region before injection, which reduces the risk of unwanted interaction between the injected beam and the CMIK.

\begin{figure}[htbp]
    \centering
    \includegraphics[width=0.95\linewidth]{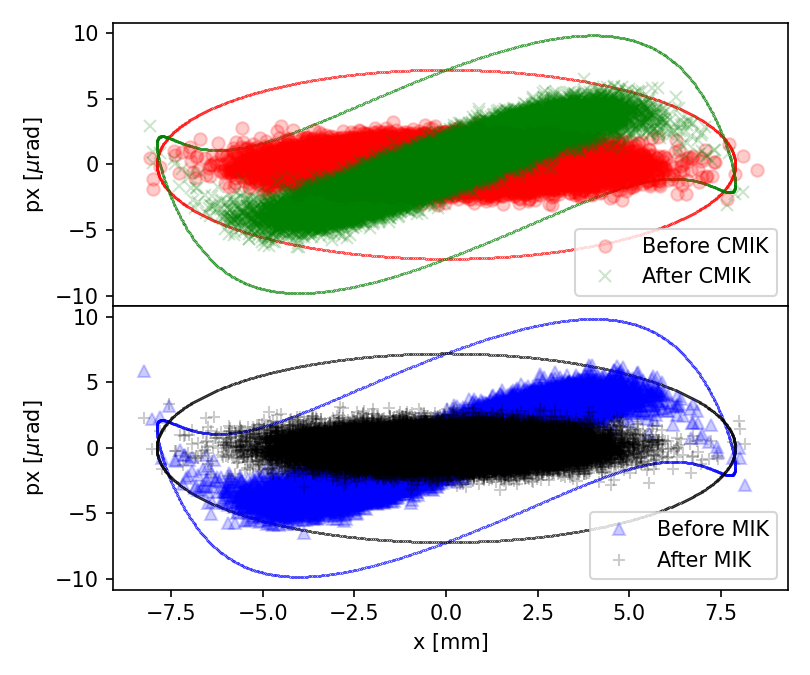}
    \caption{  
    Single-pass horizontal phase space of the circulating beam in the conceptual Z mode $\pi$ phase-advance layout. Upper: before (red) and after (green) the CMIK. Lower: before (blue) and after (black) the MIK. The ellipses correspond to $9\sigma_x$ betatron on-momentum particles.
    }
    \label{fig:pulsed_conductor_pi_z}
\end{figure}

The circulating beam compensation performance is also evaluated in the Z mode using the pulsed-conductor model.
To achieve $\pi$-phase-advance compensation, the CMIK and MIK kick functions should satisfy the corresponding relation in Table~\ref{tab:cmik_mik_relation}.
Figure~\ref{fig:pulsed_conductor_pi_z} shows the circulating beam distributions before and after the CMIK-MIK system in the horizontal phase space.

For the tracking, a Gaussian circulating beam distribution is used together with a $9\sigma$ betatron ellipse. 
The particles on the ellipse are on-momentum and therefore experience the design phase advance between the CMIK and MIK, so that the condition of Eq.~\eqref{eq:phase_condition} is exactly satisfied and their distortion is essentially fully compensated. 
Off-momentum particles sample a slightly different phase advance because of the chromaticity of the transfer section, so the compensation condition holds only approximately for them; the residual distortion nevertheless remains small across the whole energy distribution, as seen in Fig.~\ref{fig:pulsed_conductor_pi_z}.
This also indicates that the $\pi$-phase-advance scheme provides substantially more complete compensation than the near-zero phase-advance scheme under the idealized optics conditions.

Because particles near the edge of the circulating beam distribution can receive relatively large kicks, the particle trajectories between the CMIK and MIK can deviate significantly from the reference orbit.
Accurate compensation therefore requires that the nonlinear transport between the two devices be well controlled; accurate optics correction along the injection straight section is equally essential.

Under idealized phase advance and field symmetry conditions, this compensation does not require the residual field in the circulating-beam region to be particularly small; even the less ideal field-free region of the pulsed-multipole model could be well compensated.
Given its higher field flexibility, the pulsed-multipole field could therefore also be used for FCC-ee top-up injection in this scheme.

Regardless of the field model, the injected beam remains subject to the injection-region field gradient, which is mitigated by transfer-line pre-compensation as in the near-zero scheme (Sec.~\ref{subsec:injected_z_mu0}).

\section{Discussion and outlook}
\label{sec:mik_topology_discussion}

The compensation study above used the pulsed-multipole and pulsed-conductor field models.
Beyond these two concepts, several other MIK technologies and optics strategies may be relevant for FCC-ee top-up injection.
They are not designed in detail here, but are discussed in this section to place the present study in a broader context.

\subsection{Alternative MIK technologies}

Two further field topologies are considered here: eddy-current-shielded kickers and double-C-type dipole kickers.

An eddy-current-shielded kicker can provide a good field-free region around the circulating beam together with a relatively flat field in the injection region.
For FCC-ee, the main difficulty is the long flat-top: the kicker field must be held for at least one revolution period, about $\SI{304}{\mu\second}$. 
A conducting shield suppresses the field in the shielded region through induced eddy currents, which oppose changes of the field; during a flat-top these currents gradually decay, so the field slowly diffuses through the shield.
The relevant figure of merit for a flat-top is therefore the magnetic-field diffusion time through a shield of thickness $l$,
\begin{equation}
    t_{d}\simeq\mu\sigma l^{2},
    \label{eq:diffusion_time}
\end{equation}
where $\mu$ and $\sigma$ are the permeability and electrical conductivity of the shield.
Keeping the field well shielded throughout the pulse requires $t_{d}\gg t_{\mathrm{flat\text{-}top}}$. 
For copper, a shield of a few millimeters already satisfies this condition: with $l\simeq\SI{8}{mm}$ one obtains $t_{d}\simeq\SI{4.7}{ms}$, more than an order of magnitude longer than the flat-top. The same thickness follows from the usual engineering rule of dimensioning the shield in skin depths: approximating the flat-top as a half-period of a sinusoid, $f_{0}\simeq1/(2\,t_{\mathrm{flat\text{-}top}}) \approx\SI{1.6}{kHz}$, gives a copper skin depth $\delta\approx\SI{1.6}{mm}$, so that five skin depths correspond to about \SI{8}{mm}. An eddy-current-shielded solution is therefore not fundamentally excluded.

This shielding, however, comes at a cost. Such a thick shield consumes transverse aperture and limits how close the injected beam can be placed to the circulating beam, which is especially constraining given the narrow FCC-ee beam separation. Moreover, because the field diffuses slowly rather than being perfectly excluded, the field in the nominal field-free region is not exactly static: it grows slightly over the flat-top as the eddy currents decay, and the transition region evolves accordingly. Depending on the field-stability requirements during the flat-top, an even thicker shield may be needed. The eddy-current-shielded topology therefore offers a good field distribution but remains challenging for the long FCC-ee flat-top, and requires a dedicated technical and beam-dynamics study before it can be compared quantitatively with the other options.

Another possible topology is a double-C-type dipole kicker~\cite{AIBA201898}. Such a device can provide a relatively uniform field in the injection region, which helps reduce the distortion of the injected beam. However, the field-free region around the circulating beam is less favorable than that of a dedicated pulsed-conductor or shielded-field design. Therefore, compensation would be required to suppress the residual kick on the circulating beam. Its flexibility is also limited, because different FCC-ee operation modes may require different field profiles. This may require several hardware configurations or independent current settings, and hence a more detailed technical design.

Overall, the four topologies considered here are complementary, each strong in one respect but limited in others. Pulsed-multipole magnets offer the greatest flexibility, but their field distribution in both the field-free and injection regions is less favorable. Pulsed-conductor provides a good field-free region, but with a less uniform injection region and limited flexibility, since the field profile is largely fixed by the conductor geometry. The eddy-current-shielded topology gives a good field distribution in both regions, but its flexibility is limited and the long FCC-ee flat-top remains a critical constraint. The double-C-type dipole kicker provides a uniform injection region, but a less favorable field-free region and limited flexibility.

The preferred topology also depends on the compensation scheme. In the near-zero phase-advance case, where the compensation is only approximate, a small intrinsic residual field is essential, so the pulsed-conductor and the eddy-current-shielded kicker, with their good field-free region, are better suited. In the $\pi$-phase-advance case, where the compensation is nearly exact, a less ideal field-free region can be tolerated, so the more flexible pulsed-multipole magnet and the double-C-type kicker, with its uniform injection region, become advantageous.

Further optimization should address the MIK field distribution and transfer-line pre-compensation, together with the local dynamic aperture, pulse stability, impedance, aperture constraints, and machine-protection requirements.

\subsection{Optics optimization for the compensation scheme}

The optics at the CMIK and MIK---in particular the beta functions, the dispersion, and the phase advance between the two devices---are design parameters rather than fixed constraints, and they directly affect the compensation. A larger beta function reduces the phase-advance variation over the CMIK-MIK separation and thus improves the near-zero compensation, but it also enlarges the circulating beam envelope and the required field-free region; conversely, a smaller beta function reduces the injected beam size and its sensitivity to field gradients. The optimum therefore balances circulating beam compensation against the field requirements in the injected beam region, and if an exact integer-$\pi$ phase advance can be engineered, a very large beta function is not required. A dedicated optimization of the optics along the injection straight section, made self-consistent across all operation modes, is left to future work.

\section{Conclusion}
\label{sec:conclusion}

A study of on-axis top-up injection with a multipole injection kicker has been carried out for FCC-ee.
The field requirements were quantified and found to be considerably more demanding than those of typical light sources, owing to the large beta function and large dispersion at the injection point: the field-free region for the circulating beam ranges from about \SI{9.6}{mm} in the Z mode to about \SI{13.3}{mm} in the $t\bar{t}$ mode, and the injected beam lies up to about \SI{24}{mm} from the reference orbit.
To meet these requirements while limiting the perturbation of the circulating beam, a compensated MIK scheme was proposed, with a CMIK upstream of the main MIK, and the optics and field-symmetry conditions for compensation were derived for different CMIK-MIK phase advances.

Four MIK field technologies were discussed---pulsed-multipole magnets, pulsed-conductor, eddy-current-shielded kickers, and double-C-type dipole kickers---of which the pulsed-conductor is the most favorable in the near-zero phase-advance compensation scheme, owing to its good field-free region around the circulating beam.
With the pulsed-conductor model, the compensated scheme was shown to be effective: in the Z mode, the near-zero phase-advance scheme reduces the horizontal emittance growth of the circulating beam to about $7\%$ and the vertical growth to below about $2\%$.
Although the conceptual $\pi$ phase-advance scheme provided nearly exact compensation for the circulating beam, further work is required to develop a complete and realistic lattice.
The injected beam was also tracked in the Z mode: because the MIK field has a non-negligible gradient across the injected beam, a preliminary design with four nonlinear correctors in the transfer line was used to pre-deform it. 
The emittance blow-up due to the MIK field gradient was well suppressed, and no particle loss was observed over 1000 turns in the simulated 5000 particle sample.

These results show that compensated MIK injection is a promising route to on-axis top-up injection for FCC-ee, keeping the circulating beam perturbation small while meeting the demanding field requirements.
The main remaining challenge is the field gradient across the injected beam, which can be mitigated by transfer-line pre-compensation and by further optimization of the MIK field distribution.
The multi-turn tracking presented here includes synchrotron-radiation damping and quantum fluctuations but does not include beam-beam, lattice errors, and collimation, among others; confirming the overall performance will require more complete studies.

\appendix
\section{Is MIK a kicker or a septum?}
As discussed in Sec.~\ref{sec:introduction}, we consider replacing the septum with an MIK in the baseline orbit-bump injection. 
The septum and kickers are, however, fundamentally different devices. 
A septum deflects only the injected beam, with a physical blade separating it from the circulating beam; it can be either a DC or a pulsed magnet. 
A kicker, in contrast, is a pulsed magnet that acts on both beams during top-up injection.

\begin{figure}[htbp]
    \centering
    \includegraphics[width=\linewidth]{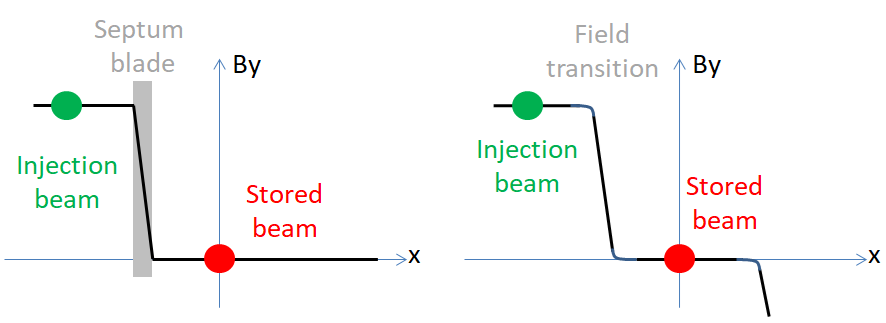}
    \caption{ 
    Vertical field $B_y$ versus horizontal position for a conventional septum magnet (left) and an MIK (right), with the injected (green) and stored (red) beams. The septum blade is replaced by a massless field transition. Idealized, for illustration only.
    }
    \label{fig:septum_vs_mik}
\end{figure}

The distinct feature of the MIK is that it deflects only the injected beam — functionally the role of a septum — with the field-transition region acting as a ``massless septum blade'' (Fig.~\ref{fig:septum_vs_mik}). 
With a conventional septum, a kicker must be installed downstream to deflect the injected beam away from the blade, which the injected beam would otherwise strike on subsequent turns. 
One might therefore argue that, if the MIK is regarded as a septum, an additional downstream kicker is still needed. This is not the case: since the ``blade'' of the MIK is massless, there is nothing for the injected beam to strike, and no downstream kicker is required. It is precisely this feature that allows the septum to be replaced by an MIK.

\bibliography{apssamp}
 
\end{document}